\documentclass{aa}
\usepackage{txfonts}
\usepackage{ifthen}
\usepackage{graphicx}

\newcommand{\Mag}{$^{m}$}

\newcommand{\mic}{$\mu$m}

\begin{document}


\title{An infrared study of galactic OH/IR stars. II. \\ The `GLMP sample' of red oxygen-rich AGB stars}

   \author{F.~M. Jim\'enez-Esteban   \inst{1,3}
        \and P. Garc\'\i a-Lario  \inst{2}
        \and D. Engels            \inst{3}
        \and J.V. Perea Calder\'on   \inst{1}
   }

\institute{ISO Data Centre / European Space Astronomy Center, Villafranca del Castillo, Apartado de Correos 50727, E-28080 Madrid, Spain.
        \and ISO Data Centre / European Space Astronomy Center, Research and Scientific Support Department of ESA, Villafranca del Castillo, Apartado de Correos 50727, E-28080 Madrid, Spain.
        \and Hamburger Sternwarte, Gojenbergsweg 112, D-21029 Hamburg, Germany.
        }

   \offprints{F.M. Jim\'enez-Esteban, \\
        \email{Francisco.Jimenez-Esteban@hs.uni-hamburg.de}}

   \date{Received 20 April 2005 / Accepted 26 July 2005}

\titlerunning{An optical/near-IR atlas of extreme OH/IR stars}

   \abstract{

We present optical and near-infrared finding charts taken from the DSS
and 2MASS surveys of 94 IRAS sources selected from the GLMP catalogue,
and accurate astrometry ($\approx$\,0.2\arcsec) for most of
them. Selection criteria were very red IRAS colours representative for
OH/IR stars with optically thick circumstellar shells and the presence
of variability according to the IRAS variability index (VAR$>$50). The
main photometric properties of the stars in this `GLMP sample' are
presented, discussed and compared with the correspondent properties of
the `Arecibo sample' of OH/IR stars studied nearlier.  We find that
37\% of the sample (N\,=\,34) has no counterpart in the 2MASS,
implying extremely high optical depths of their shells. Most of the
sources identified in the 2MASS are faint (K$\ga$8) and are of very
red colour in the near-infrared, as expected. The brightest 2MASS
counterpart (K=5.3 mag) was found for IRAS\,18299--1705. Its blue
colour H--K=1.3 suggests that IRAS\,18299--1705 is a post-AGB star.
Few GLMP sources have faint but relatively blue counterparts. They
might be misidentifed field stars or stars that recently experienced a
drop of their mass loss rates. The `GLMP sample' in general is made of
oxygen-rich AGB stars, which are highly obscured by their
circumstellar shells. They belong to the same population as the
reddest OH/IR stars in the `Arecibo sample'.


}

   \maketitle

   \section{\label{intro}Introduction}

This is the second of a series of three papers devoted to the study of
galactic OH/IR stars. In the first two papers we characterize in the
optical and in the near-infrared two large samples of OH/IR stars: the
`Arecibo sample' studied in Jim\'enez-Esteban et al.
\cite{Jimenez-Esteban05a} (hereafter Paper\,I), and the `GLMP sample'
studied here. In a forthcoming paper (Jim\'enez-Esteban et al.
\cite{Jimenez-Esteban05b}; Paper\,III) we combine both samples to
study the origin of the long-spread sequence of far-infrared colours
found by IRAS for oxygen-rich AGB stars, and propose evolutionary
links between the various classes of AGB stars and Planetary Nebulae
identified in our Galaxy.

In Paper\,I we discussed the main optical/near-infrared properties of
a well-defined sample of far-infrared selected AGB stars showing OH
maser emission. This sample, the `Arecibo sample', consists of 385
IRAS sources, which were detected in the 1612\,MHz OH maser line with
the Arecibo radio telescope (Eder et al. \cite{Eder88}; Lewis et
al. \cite{Lewis90}; Chengalur et al. \cite{Chengalur93}). The sample
was obtained from a complete survey of IRAS sources with flux
densities $\ge$\,2\,Jy at 25\,\mic, with declination
0\,$<$\,$\delta$\,$<$\,37\degr\, and appropriate colours of AGB stars
(Olnon et al. \cite{Olnon84}). The OH maser detections qualify the
IRAS sources as O-rich AGB stars. This sample is mainly constituted by
optically visible sources ($\approx$\,2/3 of the sample), having
optically thin circumstellar envelopes (CSE), and by a smaller
contribution of optically invisible ones ($\approx$\,1/3 of the
sample), having thick CSEs. These OH/IR stars are distributed over a
wide range of the near-infrared J-H\,vs.\,H-K colour-colour diagram,
which was shown in Paper\,I to be an extension toward redder colours
of the area where optically visible Mira variables are normally found
(Whitelock et al. \cite{Whitelock94}).

The `Arecibo sample' predominantly represents the bluer (and probably
less massive) population of galactic OH/IR stars. A minority of highly
obscured OH/IR star is present. To increase the sample of such OH/IR
stars we compiled a new sample, taken this time from the GLMP
catalogue (Garc\'{\i}a-Lario \cite{Garcia-Lario92}).  The purpose of
the following analysis is to determine the optical/near-infrared
properties of this sample of very red (and probably of higher mass)
AGB stars, and to establish connections with the properties of the
`Arecibo sample'. In a similar way as we did for the Arecibo sources
we have created an atlas of finding charts combining optical images
from the Second Digitized Sky Survey (DSS2; Djorgovski et
al. \cite{Djorgovski01}) and near-infrared images from the Two Micron
All Sky Survey (2MASS; Cutri et al.  \cite{Cutri03}).

Section \ref{sample} contains a description of the selected sample. In
Section \ref{ident} the process of identification of the
optical/near-infrared counterparts is explained. This is followed by a
brief description of the atlas contents. The results obtained are
discussed in Section \ref{prop}. The main conclusions are presented in
Section \ref{conc}.

   \section{\label{sample}Sample selection}

The GLMP catalogue (Garc\'{\i}a-Lario \cite{Garcia-Lario92}) contains
1084 IRAS sources with [12]--[25]\,vs.\,[25]--[60] colours similar to
those shown by planetary nebulae (PNe). To include an object in this
catalogue the source must have been well detected
(IRAS-FQual\,$\ge$\,2) in at least the three IRAS photometric bands
centred at 12, 25 and 60\,$\mu$m and obey the following constraints:

i) $F_{\nu}(12\mu m)/F_{\nu}(25\mu m)$\,$\le$\,0.50

ii) $F_{\nu}(25\mu m)/F_{\nu}(60\mu m)$\,$\ge$\,0.35 

iii) $F_{\nu}(60\mu m)/F_{\nu}(100\mu m)$\,$\ge$\,0.60, if a reliable
measurement (IRAS-FQual\,$\ge$\,2) was available in the 100\,$\mu$m
band.


\begin{table}

\caption[GLMP objects in common with the `Arecibo
sample']{\label{tab_ext:common_obj}Objects in common to the GLMP
catalogue and to the `Arecibo sample'}

\begin{center}
\begin{tabular}[t]{cccc}
\hline\hline\noalign{\smallskip}
GLMP & IRAS & GLMP & IRAS \\
\noalign{\smallskip}\hline\noalign{\smallskip}
842 & 18517+0037 & 851 & 18549+0208 \\
876 & 19065+0832 & 877 & 19067+0811 \\
882 & 19081+0322 & 901 & 19183+1148 \\
902 & 19188+1057 & 920 & 19283+1944 \\
934 & 19352+2030 & 939 & 19374+1626 \\
958 & 19565+3140 & 960 & 19576+2814 \\
966 & 20023+2855 & 972 & 20043+2653 \\
983 & 20137+2838 & 999  & 20272+3535 \\
\noalign{\smallskip}\hline
\end{tabular} 
\end{center}
\end{table}


Thus, the GLMP catalogue is made up by a heterogeneous collection of
far-infrared selected sources distributed over the full sky with very
red IRAS colours ([12]--[25]\,$\ge$\,0.75), containing PNe, but also a
considerable number of AGB and post-AGB stars (apart from a small
percentage of `contaminant' sources like T-Tauri stars, Herbig Ae-Be
stars, ultracompact H\,II regions and even a few Seyfert galaxies).
To identify the O-rich AGB stars among them, we selected objects
located in a characteristic region of the IRAS two-colour diagram
along the sequence of colours predicted for O-rich AGB stars with
increasing mass loss (Bedijn \cite{Bedijn87}; in the following we will
name this path the `O-rich AGB sequence') (see Fig.\,1) having a high
(VAR $>$ 50) IRAS variability index. They are classified in the GLMP
catalogue as `variable OH/IR stars'. Indeed, most but not all of these
stars have OH maser detections at 1612 MHz and are therefore genuine
OH/IR stars. The rest was either never observed in OH or not
detected. Nevertheless, they were classified by Garc\'{i}a-Lario as
`variable OH/IR stars', in accordance with Lewis (\cite{Lewis92}), who
found that these stars called by him `OH/IR star mimics' are indeed
O-rich variable AGB stars.

Two sources with low VAR index, namely IRAS\,11438--6330 and
IRAS\,12358--6323, were included in addition because near infrared
photometric observations taken from the literature confirmed that both
objects are strongly variable in the near-infrared (Gaylard et
al. \cite{Gaylard89}; Lepine et al. \cite{Lepine95}; Hu et
al. \cite{Hu93}; Garc\'{\i}a-Lario et al. \cite{Garcia-Lario97}) and,
thus, can also be considered as `variable OH/IR stars'.

The variability cut, while efficient in excluding non-AGB objects,
introduces a bias. Only $\approx$\,70\% of the sky was surveyed three
times during the IRAS mission, while 20\% was observed only twice, and
thus the variable sources in some parts of the sky were more likely
recognized than others. Furthermore, the variability detection
significantly depends on colour. Due to the short lifetime of the IRAS
satellite ($\approx$10\,months), it was more probable to observe
variability in OH/IR stars with shorter periods (bluer IRAS colours)
than in those with longer periods (redder IRAS colours).

The selected sources form a group of 110 OH/IR stars of which 16 are
also part of the `Arecibo sample' of OH/IR stars (see Table
\ref{tab_ext:common_obj}). Since the optical/near-infrared
characteristics of the sources in the `Arecibo sample' were already
presented and discussed in Paper\,I, we have not analysed again the
objects in common. Then, after these are excluded, the resulting `GLMP
sample' is formed by a total of 94 OH/IR stars with very red IRAS
colours.

\begin{figure*}
\begin{center}
   \includegraphics[width=13cm]{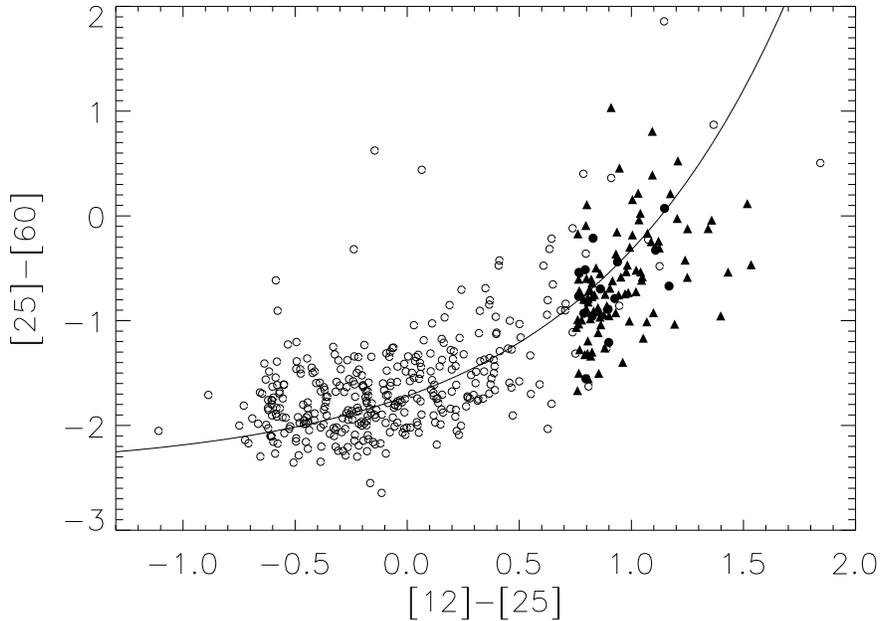} 
    \caption[IRAS colour-colour diagram for the `GLMP
             sample']{\label{fig_ext:IRAS_color}\,\, The position in
             the IRAS two-colour diagram of OH/IR stars in the `GLMP
             sample' (filled triangles) is compared with the position
             of the OH/IR stars in the `Arecibo sample' (open
             circles). Filled circles are used for the few objects in
             common. The solid line is the `O-rich AGB sequence' (see
             text), and the IRAS colours are defined as:
             [12]$-$[25]\,=\,$-$2.5\,log$\frac{F_{\nu}(12)}{F_{\nu}(25)}$
             and
             [25]$-$[60]\,=\,$-$2.5\,log$\frac{F_{\nu}(25)}{F_{\nu}(60)}$.}
\end{center} 
\end{figure*}

The position of the `GLMP sample' in the IRAS two-colour diagram is
shown in Fig.\,\ref{fig_ext:IRAS_color}, together with the `O-rich AGB
sequence', and the position of the `Arecibo sample'. The `GLMP sample'
occupies the red part in this diagram, increasing considerably the
number of objects classified as OH/IR stars or O-rich AGB stars in
this color range.


   \section{\label{ident}Identification of the optical/IR counterparts}

In order to determine the optical/near infrared counterparts of the 94
OH/IR stars in our sample we first determined the best coordinates
available from existing catalogues and then searched for plausible
counterparts at these locations both on the near-infrared images taken
from the 2MASS and on the optical images taken from the DSS2.

\subsection{Cross-correlation with the MSX} 

Improved coordinates with respect to those originally provided by the
IRAS Point Source Catalogue (accuracy typically between
10\arcsec\,--\,15\arcsec) can be obtained by cross-correlating our
sample with the MSX Point Source Catalogue (MSX6C) (Egan et
al. \cite{Egan03}), which provides coordinates of the mid-infrared
counterparts with an accuracy of $\approx$\,2\arcsec. The MSX survey
is limited to low galactic latitudes ($\le\,\mid$6$\mid^{\circ}$) and,
therefore, not all GLMP sources have an associated MSX entry. However,
the accuracy of the MSX positions is in many cases essential to
identify the near-infrared counterpart, in particular in crowded
regions along the galactic plane and/or close to the Galactic Centre,
and for extremely red objects only marginally detectable in the K
band.

We searched for the MSX counterpart in an area of 60\arcsec\, radius
around the original IRAS position. When more than one MSX source was
found, we selected the reddest MSX counterpart, because only the
reddest sources show MSX photometric data consistent with the IRAS
photometry. In almost all cases the reddest MSX counterpart was also
the nearest with respect to the search position.

Out of a total of 85 objects located within the MSX sky survey area,
we found that 82 objects (96\% of them) had a plausible mid-infrared
counterpart in the MSX6C (see Tables \ref{tab_ext:astrometry_2MASS}
and \ref{tab_ext:astrometry_MSX}). This high rate is similar to that
found for the `Arecibo sample' (Paper\,I). The 3 sources without MSX
counterpart were probably missed because they are located at the edges
of the MSX survey area, at galactic latitudes
$\mid$b$\mid$\,$>$\,5.3$^{\circ}$. For these 3 sources, plus the 9
additional ones not covered by MSX, we could not improve the original
IRAS astrometry.

\subsection{2MASS near-infrared counterparts} 
\subsubsection{General search method}

Near-infrared counterparts were searched for in the Two Micron All Sky
Survey (2MASS). We first inspected the 2MASS images and obtained the
position and the photometry of the selected counterpart mostly from
the 2MASS Point Source Catalogue (2MASS-PSC), which has an astrometric
accuracy of $\approx$\,0.2\arcsec. Depending on whether the
coordinates were from IRAS or from MSX, we inspected a circle with
radius 30\arcsec\, or 6\arcsec, respectively, centered on the nominal
source position and searched for a plausible (i.e. redder than
average) counterpart. In approximately one third of the cases only one
source was found showing redder than average colours. In another third
of the cases more than one red object was found. We always selected
the redder object, which turned out to be also the nearest with two
exceptions: IRAS\,17392--3020 and IRAS\,19087+1006.

Not all near-infrared counterparts identified were actually included
in the 2MASS-PSC. Five of them were identified directly on the 2MASS
images. These counterparts were probably not included in the PSC
because of blending with nearby sources or due to insufficient
signal-to-noise ratio. For example, close to the MSX position of
IRAS\,17151--3642 only a very bright source
(2MASS\,259.622353--36.768497) was listed in the PSC, while on the
images an additional object invisible in the J-band appears on the H-
and K-band images very close to the bright source. This very red
object was selected then as the most reliable counterpart. A similar
case is IRAS\,17495--2534. The near-infrared counterparts of
IRAS\,17428--2438 and IRAS\,18091--2437 are faint and only visible in
the K-band, probably with a flux below the detectability threshold.
Finally, IRAS\,18182--1504 is clearly in H- and K-band images, but
there is no entry in the 2MASS-PSC.  For all these sources we derived
their astrometric coordinates directly from the 2MASS images by
determining the centroid of the point-like emission associated with
the near-infrared counterpart.

A peculiar case is IRAS\,18000--2835. One 2MASS counterpart is located
only $\approx$\,1.8\arcsec\, off the MSX position, and shows the
strongest (K\,=\,5.05) near-infrared brightness in the sample,
although with relatively blue colors (H--K\,=\,0.7). A second 2MASS
counterpart was found in addition $\approx$\,3.8\arcsec\, off the MSX
position, also showing strong (K\,=\,7.14) near-infrared brightness,
but with redder colours (H--K\,=\,1.4). From the confusion flags
quoted by the 2MASS-PSC for these two sources it follows that the
photometry of both stars is contaminated. They may be a binary system,
as there is very little chance to find two such bright stars so close
together.  Because of the low spatial resolution of the IRAS
satellite, it is possible that the IRAS\,18000--2835 fluxes are a
combination of the individual fluxes of both stars. Furthermore, this
source was not detected in the 1612 MHz OH maser line by te Lintel
Hekkert (\cite{teLintelHekkert91}) which leaves its classification as
an oxygen-rich AGB star open. Therefore, we omitted IRAS\,18000--2835
from the following discussions.

65 2MASS near-infrared candidate counterparts were found by this
method among the now 81 IRAS sources with MSX identifications, while
in 16 cases (20\% of the sources with associated MSX counterparts) no
object was found. Out of the 12 objects without MSX coordinates we
were able to find suitable near-infrared candidate counterparts in ten
cases. IRAS\,18195--2804 and IRAS\,18479--2514 did not show any
plausible counterpart in the 2MASS.

\subsubsection{Constraints due to contamination by field stars \label{constraints}}


\begin{figure*}
\begin{center}
   \resizebox{\hsize}{!}{\includegraphics{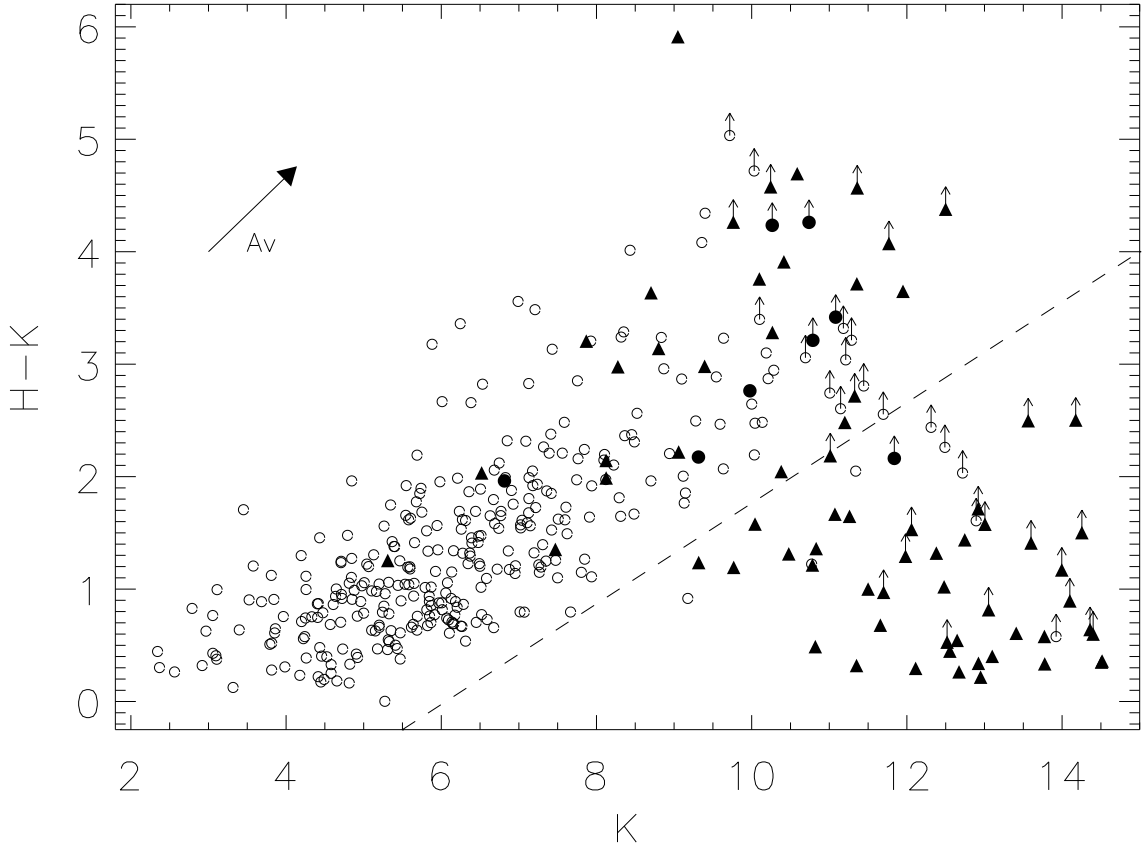}}
    \caption[H--K\,vs.\,K magnitude diagram of the `GLMP
    sample']{\label{fig_ext:KvsH-K}\,\, H--K\,vs.\,K magnitude diagram
    of all 2MASS candidate counterparts of the `GLMP sample' (filled
    triangles).  The position of the OH/IR stars in the `Arecibo
    sample' is also shown for comparison (open circles). Filled
    circles mark objects in common and H--K lower limits are
    represented by vertical arrows. The dashed line separates the
    counterparts with blue H--K colours suspected to be contaminated
    by field stars.  The reddening vector corresponds to
    A$_{V}$\,=\,10\Mag.}
\end{center}
\end{figure*}


Among the candidate counterparts a surprisingly large fraction did not
show the extremely red near-infrared colors expected for this
sample. This is shown in Figure \ref{fig_ext:KvsH-K}, where we have
plotted the near-infrared colour H--K as a function of K magnitude for
all 2MASS candidate counterparts with near-infrared photometry
available.  For comparison also the `Arecibo sample' is plotted. For
the sources not detected in H, lower limits for the H--K colour were
calculated and marked with vertical arrows.

A general correlation exists between the H--K colour and the K
magnitude for the `Arecibo sample', although there is a large scatter
in both axes. This distribution was explained in Paper I by an
increase of optical thickness of the CSEs (increase of H--K colour)
leading to an increase of obscuration in the near-infrared (decrease
in K-band brightness). The scatter is the result of i) the strong
variability with typical amplitudes
1\Mag\,$\la$\,$\Delta$K\,$\la$\,2\Mag\, rising in the most extreme
cases up to 4\Mag; and ii) the range of distances of 1--5 kpc
(Paper\,III), and iii) the variability of H--K colours which can reach
$\Delta$ (H--K) $\geq$ 1$^m$ (Harvey et al. \cite{Harvey74}; Engels et
al. \cite{Engels83}; Jim\'enez-Esteban et
al. \cite{Jimenez-Esteban05c}).

Only part of the 2MASS candidate counterparts follow the general
correlation shown by the `Arecibo sample'. They are located to the
left of the dashed line in Figure \ref{fig_ext:KvsH-K}, which was
fitted by eye to outline the lower limits of H--K colours of the
`Arecibo sample'. Hereafter this group of counterparts is referred to
as the `red subsample'.  For K\,$>$\,9 many 2MASS candidate
counterparts (as well as three of the `Arecibo sample') are
surprisingly blue with H--K\,$\la$\,2 (to the right of the dashed line
in Figure \ref{fig_ext:KvsH-K}; hereafter `blue subsample').  These
candidates do not follow the correlation of redder colours with
fainter K--band brightness.

A possible explanation for the objects of the `blue subsample' is
misidentification with field stars. Despite a relatively small search
box for the majority of the sources, the probability of
misidentification is not negligible. The stellar fields were in many
cases very crowded since some of them correspond to regions of very
low galactic latitude and/or are located in the direction of the
Galactic Center. Then, more than one 2MASS point source was usually
($\approx$\,30\%) found within the search area.

\begin{figure}
\begin{center}
   \resizebox{\hsize}{!}{\includegraphics{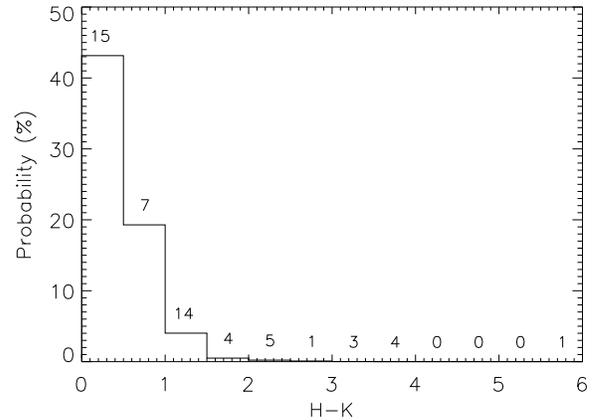}}

    \caption[]{\label{fig_ext:probab_H-K}\,\, Probability of finding a
      field star within the search area as a function of the
      near-infrared H--K colour. The rate of 2MASS candidate
      counterparts actually found is given in percentage on top of
      each bin.}

\end{center}
\end{figure}

In order to quantify the contamination we determined the probability
of misidentifications as a function of the H--K colour. We randomly
selected 9 fields in the vicinity of GLMP sources and collected all
2MASS\_PSC entries with H- and K-band detections within a radius of
$<$ 2.5\arcmin\, from their MSX coordinates. With a total number of
$\approx$\,4\,000 sources, we calculated the probability to find a
field star within a circle of 6\arcsec\, as function of the H--K
colour. The result is shown in Figure\,\ref{fig_ext:probab_H-K}.


If we now consider the whole sample of GLMP sources, we can see that
they follow a different colour distribution, showing much redder H--K
colours in general than expected from the probability analysis.  The
percentage of 2MASS candidate counterparts actually found is given on
top of each colour bin in Figure\,\ref{fig_ext:probab_H-K}. This
clearly suggests that the probability of misidentification is
negligible for the redder counterparts. The chance to find a field
star with H--K\,$\ge$\,1.5 is below 1\%, while we find that 18\% of
the 2MASS candidate counterparts show such red colours. Thus, all of
them can be considered as plausible counterparts, irrespective their
compliance with the general correlation shown by the `Arecibo sample'.

The 2MASS candidate counterparts with H--K\,$<$\,1.0 can be explained
completely as misidentifications with field stars, in accordance with
the results obtained from the `Arecibo sample'. In Paper\,I we showed
that the handful of Arecibo OH/IR stars with red IRAS colours
([12]--[25]\,$>$\,0.5) and blue near-infrared colours (H--K\,$<$\,1)
are not expected to be variable OH/IR stars, but to be candidate
post-AGB stars instead. The `GLMP sample', however, was selected
imposing that all sources must have a high IRAS variability index
(VAR\,$>$\,50), which should exclude post-AGB stars from the
sample. We therefore rejected all 16 2MASS candidate counterparts with
H--K\,$<$\,1.0 because of being most likely reddened field stars.

Finally, it remains to consider the 11 2MASS candidate counterparts
with 1.0\,$\le$\,H--K\,$<$\,1.5. The probability for misidentification
with a field star is 4\% in this colour range.  From the GLMP sources
with MSX counterpart (N\,=\,10) we expect therefor N\,$\approx$\,3
misidentified counterparts in this colour range, while 10 were
actually found. Two of them with K\,$<$\,8 follow the general
correlation shown by the `Arecibo sample', leaving $\approx$\,6
counterparts which cannot be explained as chance coincidences with
field stars.

The possible nature of the sources from the `blue subsample' will be
discussed further in Section \ref{prop}. Observations to detect
variability in these counterparts are required to confirm the
identifications.

\subsection{Consistency checks}
We made several consistency checks involving correlations between
IRAS, MSX and 2MASS positions to validate the reliability of the MSX
and 2MASS counterparts. In Table \ref{tab_ext:diff_coord_2} we list
the median and the mean separation (in arcsec) in Right Ascension
($\alpha$) and Declination ($\delta$) between 2MASS coordinates and
the IRAS and MSX ones, together with the associated standard
deviations. The median angular separation between the 2MASS and MSX
positions is 1.3\arcsec, which is similar to the median separation of
1.7\arcsec\ for the Arecibo sources found in Paper\,I.


\begin{table}

\caption[Comparison among 2MASS-IRAS-MSX coordinates]
{\label{tab_ext:diff_coord_2}Median and mean separation (and standard
deviation) between 2MASS and IRAS coordinates and between 2MASS and
MSX coordinates.}

\begin{center}
\begin{tabular}[t]{lrcccc}
\hline\noalign{\smallskip}
 & N & $\Delta\alpha$ & $\Delta\alpha$ ($\sigma_{\Delta\alpha}$) &
 $\Delta\delta$ & $\Delta\delta$ ($\sigma_{\Delta\delta}$) \\ & &
 median & mean & median & mean \\
\hline\noalign{\smallskip}\noalign{\smallskip}
2MASS--IRAS& 59 & 3.8\arcsec & 7.6\arcsec\, (8.7\arcsec) & 1.7\arcsec & 2.0\arcsec\, (2.4\arcsec) \\
2MASS--MSX & 49 & 0.9\arcsec & 1.2\arcsec\, (1.1\arcsec) & 1.0\arcsec & 1.4\arcsec\, (1.3\arcsec) \\
\noalign{\smallskip}\hline
\end{tabular} 
\end{center}
\end{table}



\begin{figure}
\begin{center}
   \resizebox{\hsize}{!}{\includegraphics{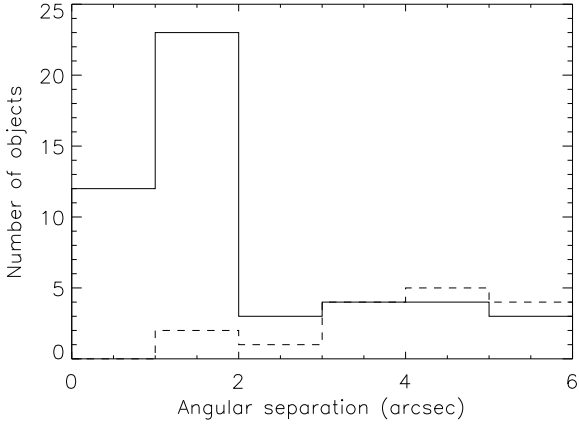}}

    \caption[Angular separation between 2MASS and MSX
    coordinates]{\label{fig_ext:MSX-2MASS}\,\, Angular separation
    between 2MASS and MSX coordinates for the 49 objects in common
    (solid line). Angular separation between 2MASS and MSX coordinates
    for the 16 blue 2MASS counterparts considered to be field stars
    (dashed line)}

\end{center}
\end{figure}


In Fig.\,\ref{fig_ext:MSX-2MASS}, we have plotted the angular
separation distribution between 2MASS and MSX coordinates.
Overplotted is the distribution of the blue 2MASS counterparts
(H--K$<$1.0) which were considered to be field stars. Their angular
separations are significantly larger, confirming their identification
as probable field stars.


\begin{figure}
   \resizebox{\hsize}{!}{\includegraphics{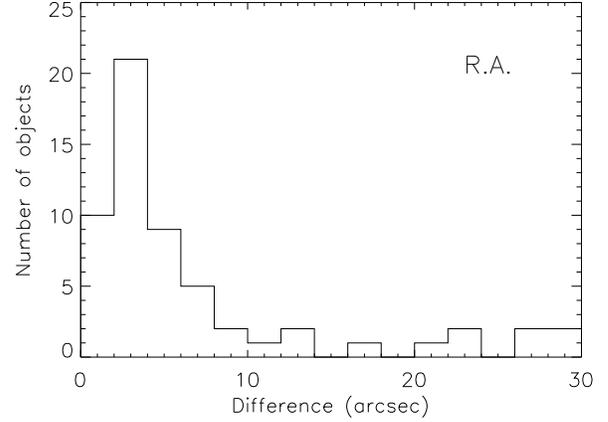}}
   \resizebox{\hsize}{!}{\includegraphics{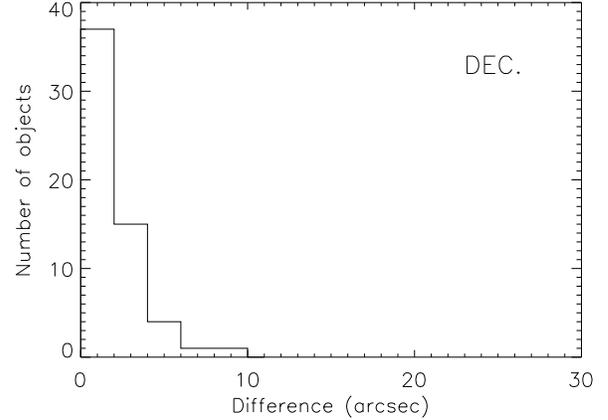}}

    \caption[$\mid$$\Delta\alpha$$\mid$ and $\mid$$\Delta\delta$$\mid$
    between 2MASS and IRAS coordinates]{\label{fig_ext:2MASS-IRAS}\,\,
    Angular separation in right ascension (upper panel) and
    declination (lower panel) between 2MASS and IRAS coordinates for
    the 59 OH/IR stars for which a near-infrared counterpart was
    found. Not included in the R.A. distribution is one source with
    $\mid$$\Delta\alpha$$\mid$\,=\,41$\arcsec$.}

\end{figure}


Much larger separations are found between 2MASS and the original IRAS
coordinates. Figure \ref{fig_ext:2MASS-IRAS} shows the distribution of
these separations. While in declination they are within 6\arcsec\,
with just two exceptions, in right ascension we find a considerable
number of objects showing separations larger than 10\arcsec. Similar
separations have been found in Paper\,I for the Arecibo sources.

Given the positional uncertainties of the IRAS-PSC of up to 1\arcmin\
(Paper I), counterparts with large 2MASS-IRAS separations might be
misidentifications, if MSX coordinates are not available.  Nine
sources with rather large 2MASS-IRAS separations in RA
($\ge$\,20\arcsec) are listed in Table \ref{tab_ext:astrometry_2MASS}.
All of them but IRAS\,18298--2111 have MSX counterparts.
IRAS\,18298--2111 is a bright IRAS source and its near-infrared
counterpart has one of the reddest colours (H--K\,=\,4.69) in our
sample. Moreover, this source is located at galactic latitude
$\mid$b$\mid$\,$=$\,5.6$^{\circ}$, where confusion is unlikely. We
conclude therefore that the identification of this counterpart is
correct.

\subsection{The final set of 2MASS counterparts}

We ended with 59 IRAS sources having a 2MASS counterpart, 49 coming
from the subset of sources with an MSX counterpart. As tentative
identifications we consider the 9 counterparts with
1.0\,$\le$\,H--K\,$<$\,1.5 located outside H--K vs. K distribution of
the Arecibo sources, and the counterparts detected only in the K-band
and with a lower limit of H--K\,$<$\,1.5, because the H--K constraint
could not be determined. The same applies for IRAS\,14562--5637,
IRAS\,17004--4119, IRAS\,17276--2846 and IRAS\,17392-3020, whose
photometry is affected by blending problems, and for
IRAS\,17495--2534, which was just barely detected both in the K and H
filters. Detection of variability is also required in these cases to
confirm the counterpart.


\begin{table*}
\caption[2MASS astrometry for the `GLMP
        sample']{\label{tab_ext:astrometry_2MASS} 2MASS and MSX counterparts of
        oxygen-rich AGB stars in the `GLMP sample'.}
\begin{center}
\begin{tabular}{clrlrrrrcc}
\hline\hline\noalign{\smallskip}
  &      & \multicolumn{2}{c}{2MASS}       & \multicolumn{2}{c}{IRAS--2MASS} & \multicolumn{2}{c}{IRAS--MSX}   &          &     \\
GLMP & IRAS & \multicolumn{2}{c}{Coordinates} & $\Delta\alpha$ & $\Delta\delta$ & $\Delta\alpha$ & $\Delta\delta$ & MSX6C\_G & Opt \\
  &      & \multicolumn{2}{c}{(J2000)}     & [\arcsec]      & [\arcsec]      & [\arcsec]      & [\arcsec]      &          &     \\
\noalign{\smallskip}\hline\noalign{\smallskip}
228 & \object{08425$-$5116} * & 08 44 04.36 & $-$51 27 41.9 &     3 &     0 &     2 &  $-$1 & 269.4419$-$05.4504 &    \\
309 & \object{11438$-$6330}   & 11 46 14.00 & $-$63 47 14.2 &  $-$2 &     2 &  $-$2 &     3 & 295.8065$-$01.8214 &    \\
320 & \object{12158$-$6443} * & 12 18 35.48 & $-$65 00 14.2 &     2 &  $-$3 &     3 &  $-$3 & 299.4688$-$02.3533 &    \\
333 & \object{12358$-$6323}   & 12 38 48.35 & $-$63 39 39.4 &  $-$1 &     1 &     0 &     0 & 301.5309$-$00.8231 &    \\
382 & \object{14247$-$6148}   & 14 28 30.73 & $-$62 01 35.7 &     0 &     0 &     3 &     0 & 314.0494$-$01.2684 &    \\
398 & \object{14562$-$5637} * & 15 00 02.39 & $-$56 49 44.2 &     6 &  $-$3 &     7 &  $-$2 & 319.8669+01.7534   &    \\
430 & \object{15471$-$4634}   & 15 50 40.33 & $-$46 43 16.9 &     2 &  $-$1 &  $-$2 &  $-$3 & 331.9853+05.7905   &    \\
435 & \object{15514$-$5323}   & 15 55 21.13 & $-$53 32 44.6 & $-$26 &     8 & $-$22 &    10 & 328.2246+00.0418   &    \\
447 & \object{16040$-$4708} * & 16 07 39.47 & $-$47 16 26.3 &     3 &  $-$2 &     3 &  $-$2 & 333.8253+03.4834   &    \\
463 & \object{16219$-$4823}   & 16 25 41.89 & $-$48 30 33.9 &  $-$1 &     0 &     0 &  $-$1 & 335.1416+00.4939   &    \\
503 & \object{16582$-$3059}   & 17 01 29.83 & $-$31 04 19.0 &  $-$6 &     0 &       &       &                    &    \\
508 & \object{17004$-$4119} * & 17 03 55.97 & $-$41 24 00.8 &     1 &     2 &  $-$2 &     0 & 344.9293+00.0136   & ** \\
517 & \object{17030$-$3053}   & 17 06 14.07 & $-$30 57 38.3 &  $-$2 &     0 &  $-$3 &     2 & 353.5472+05.9448   &    \\
532 & \object{17107$-$3330}   & 17 14 05.06 & $-$33 33 58.0 &  $-$2 &  $-$1 &  $-$4 &  $-$2 & 352.4216+03.0672   &    \\
541 & \object{17151$-$3642} * & 17 18 29.74 & $-$36 46 07.2 &  $-$5 &     5 &  $-$4 &     0 & 350.3340+00.4777   &    \\
548 & \object{17171$-$2955}   & 17 20 20.25 & $-$29 58 22.8 &  $-$3 &     1 &  $-$3 &  $-$1 & 356.1310+04.0515   & ** \\
564 & \object{17242$-$3859} * & 17 27 38.65 & $-$39 02 10.1 &    12 &  $-$2 &    12 &  $-$1 & 349.4829$-$02.2940 & ** \\
573 & \object{17276$-$2846} * & 17 30 48.35 & $-$28 49 01.7 &     3 &     1 &     5 &  $-$2 & 358.3668+02.8054   &    \\
582 & \object{17316$-$3523}   & 17 34 57.50 & $-$35 25 52.6 &     6 &     0 &     6 &     1 & 353.2979$-$01.5367 &    \\
583 & \object{17317$-$3331}   & 17 35 02.72 & $-$33 33 29.4 &  $-$7 &  $-$1 &  $-$5 &     0 & 354.8833$-$00.5384 &    \\
589 & \object{17341$-$3529}   & 17 37 30.29 & $-$35 31 04.7 & $-$17 &     3 & $-$17 &     4 & 353.5039$-$02.0204 &    \\
594 & \object{17350$-$2413}   & 17 38 08.85 & $-$24 14 49.3 &  $-$2 &     0 &  $-$3 &     0 & 003.1080+03.8899   & ** \\
595 & \object{17351$-$3429} * & 17 38 26.29 & $-$34 30 41.0 &  $-$2 &  $-$1 &  $-$4 &  $-$2 & 354.4579$-$01.6437 &    \\
604 & \object{17367$-$3633} * & 17 40 07.63 & $-$36 34 41.4 &  $-$4 &  $-$3 &  $-$2 &  $-$2 & 352.8880$-$03.0333 &    \\
619 & \object{17392$-$3020} * & 17 42 30.81 & $-$30 22 07.3 & $-$22 &  $-$2 & $-$17 &     0 & 358.4249$-$00.1744 &    \\
626 & \object{17417$-$1630}   & 17 44 39.87 & $-$16 31 42.1 &  $-$4 &  $-$2 &       &       &                    &    \\
636 & \object{17428$-$2438} * & 17 45 56.97 & $-$24 39 57.8 &  $-$2 &  $-$2 &  $-$3 &  $-$2 & 003.6854+02.1592   &    \\
649 & \object{17467$-$4256}   & 17 50 20.50 & $-$42 57 18.5 &  $-$3 &  $-$1 &       &       &                    &    \\
656 & \object{17495$-$2534} * & 17 52 39.59 & $-$25 34 39.5 &     0 &     0 &     0 &  $-$2 & 003.6848+00.3883   &    \\
659 & \object{17504$-$3312}   & 17 53 50.24 & $-$33 13 26.3 & $-$41 &     2 & $-$41 &     0 & 357.2197$-$03.7085 &    \\
672 & \object{17545$-$3317}   & 17 57 49.20 & $-$33 17 47.6 &     1 &  $-$1 &     0 &  $-$2 & 357.5737$-$04.4663 &    \\
685 & \object{17577$-$1519}   & 18 00 36.12 & $-$15 19 44.6 & $-$13 &  $-$2 & $-$15 &  $-$2 & 013.4930+03.9153   &    \\
691 & \object{17583$-$3346}   & 18 01 39.27 & $-$33 46 00.5 &  $-$5 &  $-$1 &       &       &                    & ** \\
704 & \object{18006$-$1734} * & 18 03 36.86 & $-$17 34 00.4 &  $-$4 &  $-$1 &  $-$4 &  $-$1 & 011.8974+02.1862   &    \\
709 & \object{18015$-$1608}   & 18 04 28.28 & $-$16 07 52.3 &  $-$8 &     0 & $-$10 &     0 & 013.2508+02.7102   &    \\
727 & \object{18081$-$0338} * & 18 10 49.48 & $-$03 38 14.2 &  $-$6 &     0 &       &       &                    &    \\
731 & \object{18091$-$0855} * & 18 11 56.55 & $-$08 54 46.1 &  $-$4 & $-$3  &  $-$3 & $-$3  &  020.4676+04.5958  & ** \\
730 & \object{18091$-$2437} * & 18 12 16.13 & $-$24 36 42.9 &  $-$6 &  $-$2 &  $-$5 &  $-$1 & 006.7152$-$02.9972 &    \\
743 & \object{18107$-$0710}   & 18 13 30.01 & $-$07 09 48.2 &  $-$3 &  $-$1 &  $-$4 &  $-$2 & 022.1975+05.0855   &    \\
755 & \object{18152$-$0919} * & 18 17 58.57 & $-$09 18 31.5 & $-$20 &  $-$1 & $-$21 &  $-$1 & 020.8254+03.0967   &    \\
756 & \object{18182$-$1504} * & 18 21 06.83 & $-$15 03 20.0 & $-$12 &  $-$4 & $-$13 &  $-$2 & 016.1169$-$00.2903 &    \\
760 & \object{18187$-$0208} * & 18 21 18.76 & $-$02 07 01.1 &  $-$4 &  $-$2 &  $-$3 &  $-$4 & 027.5923+05.7240   &    \\
767 & \object{18201$-$2549}   & 18 23 12.28 & $-$25 47 58.7 &  $-$1 &     2 &       &       &                    &    \\
768 & \object{18211$-$1712}   & 18 24 05.54 & $-$17 11 11.9 &  $-$6 &  $-$1 &  $-$7 &  $-$1 & 014.5704$-$01.9215 &    \\
784 & \object{18257$-$1052}   & 18 28 30.97 & $-$10 50 48.5 &     3 &  $-$6 &     4 &  $-$2 & 021.4566+00.4911   &    \\
786 & \object{18262$-$0735}   & 18 28 59.86 & $-$07 33 23.0 &  $-$5 &  $-$2 &  $-$6 &  $-$3 & 023.6510+01.5062   &    \\
787 & \object{18266$-$1239}   & 18 29 28.74 & $-$12 37 53.1 &     8 &     1 &   9   &     3 & 019.2086$-$00.9527 & ** \\
791 & \object{18277$-$1059}   & 18 30 30.92 & $-$10 57 33.4 &    28 &     1 &    29 &     1 & 020.8085$-$00.4024 &    \\
793 & \object{18298$-$2111}   & 18 32 48.56 & $-$21 09 37.6 &    28 &     1 &       &       &                    &    \\
794 & \object{18299$-$1705}   & 18 32 50.75 & $-$17 02 48.5 &  $-$4 &     0 &  $-$4 &  $-$1 & 015.6638$-$03.7108 & ** \\
799 & \object{18310$-$2834}   & 18 34 13.78 & $-$28 32 21.6 &  $-$2 &  $-$2 &       &       &                    &    \\
810 & \object{18361$-$0647}   & 18 38 50.54 & $-$06 44 49.6 & $-$32 &    14 & $-$34 &    14 & 025.4948$-$00.2879 &    \\
826 & \object{18432$-$0149} * & 18 45 52.40 & $-$01 46 42.5 &     6 &     0 &     6 &     0 & 030.7147+00.4264   &    \\
831 & \object{18460$-$0254}   & 18 48 41.97 & $-$02 50 25.4 & $-$23 & $-$10 & $-$24 &  $-$6 & 030.0908$-$00.6866 &    \\
835 & \object{18475$-$1428}   & 18 50 22.90 & $-$14 24 30.7 &  $-$6 &  $-$3 &       &       &                    &    \\
839 & \object{18488$-$0107} * & 18 51 26.02 & $-$01 03 56.0 &  $-$3 &     2 &  $-$6 &  $-$2 & 031.9844$-$00.4849 &    \\
878 & \object{19069+1335}     & 19 09 16.61 &   +13 40 27.9 &  $-$6 &  $-$3 &  $-$6 &  $-$3 & 047.1168+02.3242   &    \\
884 & \object{19087+1006}   * & 19 11 10.00 &   +10 11 43.7 & $-$30 & $-$14 & $-$28 &  $-$8 &  044.2404+00.3090  &    \\
892 & \object{19122$-$0230}   & 19 14 54.84 & $-$02 25 28.8 &  $-$7 &  $-$2 &       &       &                    &    \\
\noalign{\smallskip}\hline   
\end{tabular}
\end{center}
{\em Note:} * tentative 2MASS identification; ** shows optical counterpart
\end{table*}


Thus, the identification strategy provided in 36 cases a plausible
near-infrared counterpart, while in 23 cases only a tentative
near-infrared counterpart is provided. These 59 sources are listed in
Table \ref{tab_ext:astrometry_2MASS}. The Table includes the IRAS
name, the astrometric position from the 2MASS-PSC, the difference to
the positions provided originally by IRAS and with the ones derived
from the MSX6C, and the MSX name, when available. The objects with
tentative identification have been labeled with an asterisk.

All the objects without an identified 2MASS counterpart but with an
MSX one are given in Table \ref{tab_ext:astrometry_MSX}. This Table
includes the IRAS name, the MSX coordinates, the difference to the
original IRAS coordinates and the associated MSX object name. Only
IRAS\,18479--2514 and IRAS\,18195-2804 had neither an MSX nor 2MASS
counterpart. For these two sources no improvement to the IRAS
coordinates was possible. Their IRAS coordinates have also been
tabulated in Table \ref{tab_ext:astrometry_MSX}.


\begin{table*}

\caption[MSX astrometry for the `GLMP
sample']{\label{tab_ext:astrometry_MSX}MSX counterparts of
        oxygen-rich AGB stars in the `GLMP sample', 
        without near-infrared counterparts identified.}

\begin{center}
\begin{tabular}{clrlrrccrr}
\hline\noalign{\smallskip}
     &      & \multicolumn{2}{c}{MSX}         & \multicolumn{2}{c}{IRAS$-$MSX}  &          \\
GLMP & IRAS & \multicolumn{2}{c}{Coordinates} & $\Delta\alpha$ & $\Delta\delta$ & MSX6C\_G \\
     &      & \multicolumn{2}{c}{(J2000)}     & [\arcsec]      & [\arcsec]      &          \\
\noalign{\smallskip}\hline\noalign{\smallskip}
239 & \object{09024$-$5019} & 09 04 03.3 & $-$50 31 32 &   0 &   4 & 270.7424$-$02.4384 \\
410 & \object{15198$-$5625} & 15 23 42.7 & $-$56 36 08 &   1 &   1 & 322.7729+00.2860 \\
418 & \object{15327$-$5400} & 15 36 35.9 & $-$54 10 29 &  $-$5 &   1 & 325.6578+01.2432 \\
465 & \object{16236$-$5332} & 16 27 39.3 & $-$53 39 04 & $-$13 &   2 & 331.6587$-$03.3024 \\
519 & \object{17055$-$3753} & 17 08 57.3 & $-$37 56 50 &   6 &   0 & 348.2670+01.3225 \\
536 & \object{17128$-$3528} & 17 16 12.9 & $-$35 32 12 &  $-$1 &   0 & 351.0725+01.5639 \\
555 & \object{17207$-$3632} & 17 24 07.4 & $-$36 35 39 &   5 &  $-$3 & 351.1189$-$00.3517 \\
566 & \object{17251$-$2821} & 17 28 18.5 & $-$28 23 57 &   3 &  $-$4 & 358.4140+03.4923 \\
577 & \object{17292$-$2727} & 17 32 23.4 & $-$27 29 58 &   4 &  $-$3 & 359.6629+03.2315 \\
585 & \object{17323$-$2424} & 17 35 26.0 & $-$24 26 32 &  $-$6 &   2 & 002.6111+04.3084 \\
602 & \object{17361$-$2358} & 17 39 15.0 & $-$23 59 55 &   3 &  $-$3 & 003.4525+03.8089 \\
603 & \object{17367$-$2722} & 17 39 52.4 & $-$27 23 32 &  $-$4 &  $-$4 & 000.6467+01.8893 \\
605 & \object{17368$-$3515} & 17 40 12.9 & $-$35 16 41 &   2 &  $-$3 & 354.0019$-$02.3600 \\
624 & \object{17411$-$3154} & 17 44 24.0 & $-$31 55 40 & $-$17 &   1 & 357.3108$-$01.3371 \\
629 & \object{17418$-$2713} & 17 44 58.8 & $-$27 14 43 &  $-$1 &  $-$2 & 001.3690+01.0025 \\
664 & \object{17521$-$2938} & 17 55 21.8 & $-$29 39 13 &   0 &   1 & 000.4720$-$02.1918 \\
673 & \object{17545$-$3056} & 17 57 48.4 & $-$30 56 25 &   0 &  $-$2 & 359.6212$-$03.2926 \\
692 & \object{17584$-$3147} & 18 01 41.8 & $-$31 47 56 &  $-$4 &  $-$2 & 359.2842$-$04.4385 \\
699 & \object{18000$-$2835} & 18 03 13.0 & $-$28 35 43 &  $-$29 &   2 & 002.2442-03.1547 \\
705 & \object{18007$-$1841} & 18 03 39.0 & $-$18 41 10 &  21 &   4 & 010.9257+01.6294 \\
712 & \object{18019$-$3121} & 18 05 12.1 & $-$31 21 45 &  $-$1 &   2 & 000.0313$-$04.8783 \\
717 & \object{18040$-$2726} & 18 07 09.0 & $-$27 25 52 &  $-$5 &  $-$4 & 003.6860$-$03.3459 \\
718 & \object{18040$-$2953} & 18 07 18.3 & $-$29 53 10 &  $-$9 &  $-$3 & 001.5478$-$04.5617 \\
732 & \object{18092$-$2347} & 18 12 20.5 & $-$23 46 57 & $-$17 &  $-$3 & 007.4526$-$02.6149 \\
733 & \object{18092$-$2508} & 18 12 21.9 & $-$25 07 21 &  $-$4 &  $-$1 & 006.2765$-$03.2604 \\
740 & \object{18100$-$1915} & 18 13 03.2 & $-$19 14 19 &  $-$6 &  $-$1 & 011.5218$-$00.5826 \\
763 & \object{18195$-$2804} * & 18 22 40.2 & $-$28 03 08 &       &       &                    \\
765 & \object{18198$-$1249} & 18 22 43.1 & $-$12 47 42 & $-$12 &  $-$2 & 018.2955+00.4291 \\
783 & \object{18257$-$1000} & 18 28 31.0 & $-$09 58 15 & $-$35 &  $-$3 & 020.6796+00.0841 \\
788 & \object{18268$-$1117} & 18 29 35.6 & $-$11 15 54 &   3 &   0 & 020.4328$-$00.3439 \\
811 & \object{18361$-$0036} & 18 38 45.1 & $-$00 33 22 &   2 &   1 & 030.9892+02.5686 \\
837 & \object{18479$-$2514} * & 18 51 00.9 & $-$25 11 06 &       &       &                    \\
843 & \object{18518+0558}   & 18 54 17.4 & +06 02 34 &  $-$1 &  $-$2 & 038.6366+02.1198 \\
888 & \object{19112+1220}   & 19 13 37.4 & +12 25 38 & $-$16 &  $-$8 & 046.4992+00.8092 \\
916 & \object{19254+1724}   & 19 27 41.1 & +17 30 36 &   1 &  $-$1 & 052.5814+00.2014 \\
\noalign{\smallskip}\hline
\end{tabular}
\end{center}
{\em Note:} * IRAS coordinates.
\end{table*}


In Table \ref{tab_ext:photometry_2MASS} we present the J, H and K
near-infrared 2MASS magnitudes (or lower limits) for the 59 sources
with a plausible or tentative near-infrared identification. The lower
limits have been obtained from nearby stars with similar or even
greater brightnesses. For the 5 objects which have been identified
only on the 2MASS images, but have no associated entry in the
2MASS-PSC, we have obtained their photometry directly from the 2MASS
images, with the exception of IRAS\,17495--2534 because it was
strongly blended. We have also measured the photometry in the K band
for IRAS\,17171--2955, as the 2MASS-PSC only provides a lower limit.


\begin{table*}

\caption[2MASS photometry for the `GLMP
sample']{\label{tab_ext:photometry_2MASS} 2MASS JHK band photometry of
the oxygen-rich AGB stars of the `GLMP sample'}

\begin{center}
\begin{tabular}{crrr}
\hline\hline\noalign{\smallskip}
IRAS &  \multicolumn{3}{c}{2MASS photometry [mag]} \\
     & J & H & K   \\
\noalign{\smallskip}\hline\noalign{\smallskip}
08425$-$5116 & 13.79 $\pm$ 0.03 & 12.19 $\pm$ 0.03 & 10.83 $\pm$ 0.02 \\
11438$-$6330 & $>$ 17  -------- & 14.96 $\pm$ 0.09 &  9.05 $\pm$ 0.02 \\
12158$-$6443 & 12.77 $\pm$ 0.02 & 10.96 $\pm$ 0.02 &  9.77 $\pm$ 0.02 \\
12358$-$6323 & $>$ 14  -------- & $>$ 13 -------- & 12.06 $\pm$ 0.04 \\
14247$-$6148 & 13.34 $\pm$ 0.03 & 10.27 $\pm$ 0.02 &  8.12 $\pm$ 0.02 \\
14562$-$5637 & 15.20 $\pm$ 0.07 & * 13.73  -------- & * 13.20  -------- \\
15471$-$4634 & $>$ 16 -------- & 15.60 $\pm$ 0.16 & 11.95 $\pm$ 0.03 \\
15514$-$5323 & 15.11 $\pm$ 0.04 & 11.62 $\pm$ 0.03 & 10.04 $\pm$ 0.02 \\
16040$-$4708 & 15.87 $\pm$ 0.09 & 14.18 $\pm$ 0.09 & 12.75 $\pm$ 0.04 \\
16219$-$4823 & $>$ 15 ------- & $>$ 14 -------- & 11.33 $\pm$ 0.04 \\
16582$-$3059 & $>$ 17 -------- & $>$ 16 -------- & 13.56 $\pm$ 0.05 \\
17004$-$4119 & 15.89 $\pm$ 0.10 & 13.23 $\pm$ 0.05 & * 11.52  -------- \\
17030$-$3053 & $>$ 17 -------- & $>$ 16 -------- & 12.50 $\pm$ 0.03 \\
17107$-$3330 & $>$ 16  -------- & 13.86 $\pm$ 0.05 & 10.10 $\pm$ 0.03 \\
17151$-$3642 & $>$ 13  -------- & ** 12.5 $\pm$ 0.5 & ** 11.5 $\pm$ 0.5 \\
17171$-$2955 & 14.71 $\pm$ 0.08 & 13.68 $\pm$ 0.09 & ** 11.2 $\pm$ 0.1  \\
17242$-$3859 & $>$ 13  -------- & $>$ 12  -------- & 11.70 $\pm$ 0.04 \\
17276$-$2846 & 10.89 $\pm$ 0.04 & * 9.47  -------- & * 8.79  -------- \\
17316$-$3523 & $>$ 16  -------- & $>$ 14  -------- & 13.01 $\pm$ 0.06 \\
17317$-$3331 & $>$ 16  -------- & $>$ 14  -------- & 10.24 $\pm$ 0.03 \\
17341$-$3529 & 14.62 $\pm$ 0.06 & 11.28 $\pm$ 0.03 &  9.06 $\pm$ 0.02 \\
17350$-$2413 & $>$ 14  -------- & $>$ 13  -------- & 11.01 $\pm$ 0.03 \\
17351$-$3429 & $>$ 14  -------- & $>$ 13  -------- & 11.98 $\pm$ 0.04 \\
17367$-$3633 & $>$ 14  -------- & $>$ 13  -------- & 12.52 $\pm$ 0.03 \\
17392$-$3020 & 14.02 $\pm$ 0.06 & 11.60 $\pm$ 0.05 & *  9.12  -------- \\
17417$-$1630 & $>$ 17  -------- & 14.32 $\pm$ 0.04 & 10.41 $\pm$ 0.02 \\
17428$-$2438 & $>$ 15  -------- & $>$ 15  -------- & ** 14.4 $\pm$ 0.2 \\
17467$-$4256 & $>$ 17  -------- & 15.07 $\pm$ 0.08 & 11.35 $\pm$ 0.02 \\
17495$-$2534 & $>$ 13  -------- & $>$ 12  -------- & $>$ 11  -------- \\
17504$-$3312 & $>$ 15  -------- & 12.91 $\pm$ 0.02 & 11.26 $\pm$ 0.02 \\
17545$-$3317 & $>$ 15  -------- & $>$ 14  -------- & 12.92 $\pm$ 0.07 \\
17577$-$1519 & 14.41 $\pm$ 0.05 & 12.43 $\pm$ 0.03 & 10.38 $\pm$ 0.02 \\
17583$-$3346 & $>$ 15  -------- & 13.55 $\pm$ 0.05 & 10.27 $\pm$ 0.02 \\
18006$-$1734 & 13.30 $\pm$ 0.04 & 11.79 $\pm$ 0.03 & 10.48 $\pm$ 0.03 \\
18015$-$1608 & $>$ 14  -------- & 12.38 $\pm$ 0.04 &  9.39 $\pm$ 0.02 \\
18081$-$0338 & 12.41 $\pm$ 0.03 & 10.55 $\pm$ 0.03 &  9.32 $\pm$ 0.02 \\
18091$-$0855 & 14.37 $\pm$ 0.04 & 13.50 $\pm$ 0.04 & 12.48 $\pm$ 0.03 \\
18091$-$2437 & $>$ 15  -------- & $>$ 15  -------- & ** 14.36 $\pm$ 0.08 \\
18107$-$0710 & 15.60 $\pm$ 0.08 & 11.07 $\pm$ 0.03 &  7.87 $\pm$ 0.02 \\
18152$-$0919 & $>$ 16  -------- & $>$ 15  -------- & 14.00 $\pm$ 0.08 \\
18182$-$1504 & $>$ 15  -------- & ** 13.70 $\pm$ 0.07 & ** 12.38 $\pm$ 0.04 \\
18187$-$0208 & 13.61 $\pm$ 0.03 & 12.00 $\pm$ 0.02 & 10.78 $\pm$ 0.02 \\
18201$-$2549 & 15.39 $\pm$ 0.06 & 11.25 $\pm$ 0.03 &  8.28 $\pm$ 0.02 \\
18211$-$1712 & $>$ 12  -------- & 11.94 $\pm$ 0.11 &  8.80 $\pm$ 0.09 \\
18257$-$1052 & 11.70 $\pm$ 0.03 &  8.82 $\pm$ 0.04 &  7.47 $\pm$ 0.03 \\
18262$-$0735 & $>$ 16  -------- & 12.34 $\pm$ 0.03 &  8.70 $\pm$ 0.02 \\
18266$-$1239 & 14.05 $\pm$ 0.02 & 12.74 $\pm$ 0.03 & 11.07 $\pm$ 0.02 \\
18277$-$1059 & $>$ 16  -------- & $>$ 14  -------- &  9.76 $\pm$ 0.03 \\
18298$-$2111 & $>$ 16  -------- & 15.28 $\pm$ 0.09 & 10.59 $\pm$ 0.02 \\
18299$-$1705 &  8.23 $\pm$ 0.02 &  6.56 $\pm$ 0.03 &  5.31 $\pm$ 0.02 \\
18310$-$2834 & $>$ 16  -------- & $>$ 15  -------- & 11.77 $\pm$ 0.03 \\
18361$-$0647 & $>$ 16  -------- & $>$ 15  -------- & 11.36 $\pm$ 0.03 \\
18432$-$0149 & $>$ 16  -------- & $>$ 15  -------- & 13.60 $\pm$ 0.05 \\
18460$-$0254 & $>$ 17  -------- & $>$ 15  -------- & 14.26 $\pm$ 0.09 \\
18475$-$1428 & 11.52 $\pm$ 0.02 &  8.55 $\pm$ 0.04 &  6.52 $\pm$ 0.02 \\
18488$-$0107 & $>$ 16  -------- & $>$ 13  -------- & 13.05 $\pm$ 0.06 \\
19069$+$1335 & $>$ 17  -------- & $>$ 16  -------- & 14.18 $\pm$ 0.08 \\
19087$+$1006 & $>$ 17  -------- & $>$ 15  -------- & 14.10 $\pm$ 0.08 \\
19122$-$0230 & 12.94 $\pm$ 0.03 & 10.11 $\pm$ 0.03 &  8.12 $\pm$ 0.03 \\
\noalign{\smallskip}\hline
\end{tabular}
\end{center}
{\em Note:} * The accuracy of the photometry is affected by deblending problems acording to 2MASS catalogue; ** Photometry obtained from the 2MASS images.
\end{table*}


\subsection{Search for optical counterparts} 

In order to search for the optical counterparts of those GLMP sources
for which we found a near-infrared counterpart, we inspected their
positions on the optical images extracted from the DSS2 (Djorgovski et
al. \cite{Djorgovski01}) in the red filter, which covers the spectral
range 6\,000\,--\,7\,000\,\AA\, with a maximum efficiency around
6\,700\,\AA. We used a 1\arcsec\,$\times$\,1\arcsec\ box centered on
the nominal 2MASS source position.

The majority (91\%) of the objects did not show any optical
counterpart, and only 8 objects had a faint optical counterpart. Based
on our expectation that the sources in the `GLMP sample' would have
optically thick CSE the association with optically visible
counterparts is questionable. However, also in the `Arecibo sample'
restricted to the same IRAS colour range ($F_{\nu}(12\mu
m)/F_{\nu}(25\mu m)$\,$\le$\,0.50) a similar rate of 90\% was found
without the optical counterpart. The identified counterparts could be
either real and possibly observed close to maximum light, or
misidentifications with reddened field stars. Variability studies of
the optical counterparts would help with their confirmation. The
objects with an optical counterpart have been labeled in Table
\ref{tab_ext:astrometry_2MASS}.

\subsection{\label{ext_sect_atlas}The atlas of optical and near-infrared counterparts}


\begin{figure*}
\begin{center}
        \includegraphics[width=12.5cm]{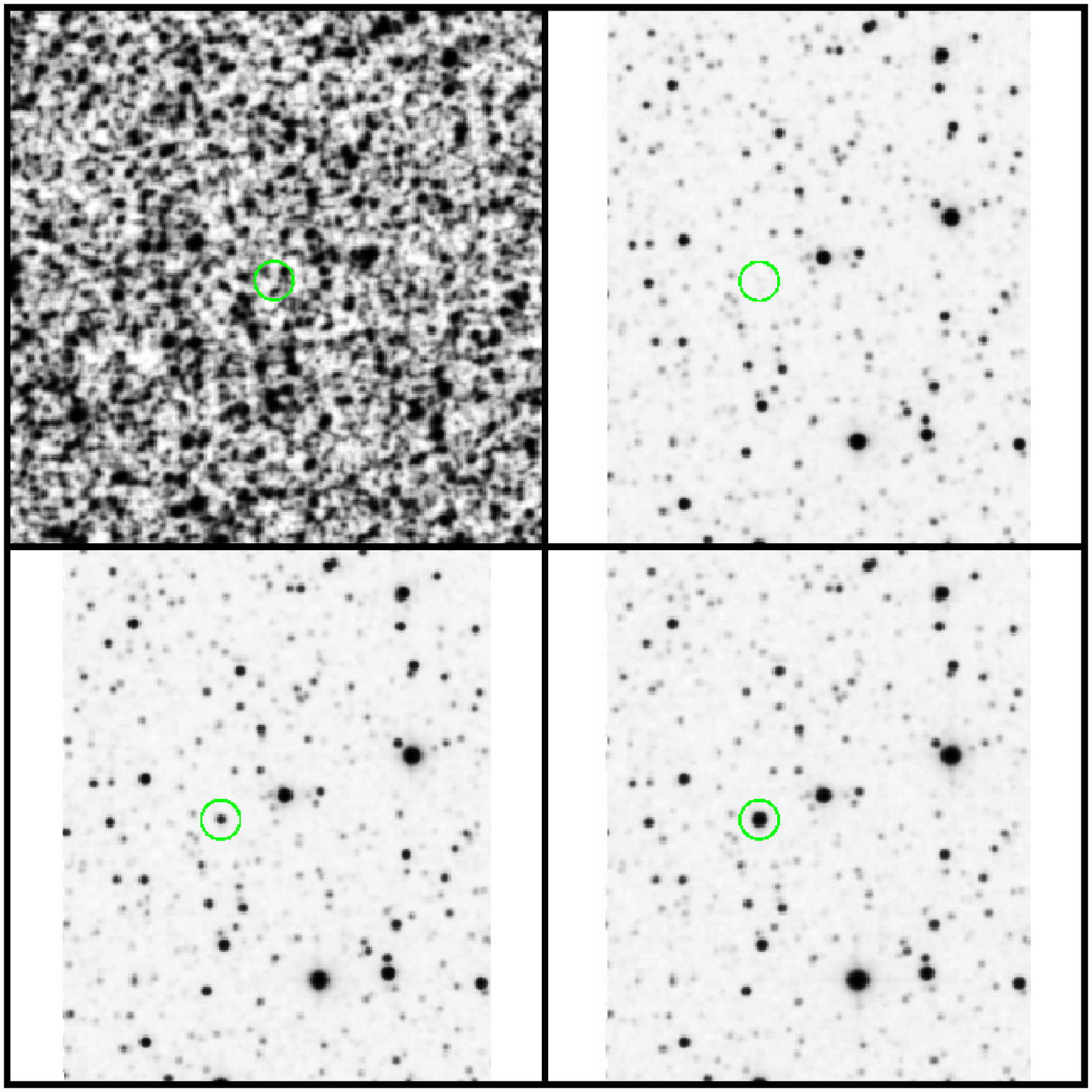} 
        \caption[Atlas images for
        IRAS\,18201--2549]{\label{fig_ext:atlas_example_2} \,\, Atlas
        images for IRAS\,18201--2549. The upper panel shows the
        optical DSS2 R band (left) and the 2MASS J band (right)
        images, and the lower panel shows the H-band (left) and K-band
        (right) images. The adopted counterpart is indicated by a
        circle.}
\end{center}  
\end{figure*}


The results of the identifications are displayed as an atlas of
finding charts. An example of the atlas images is given in
Fig.\,\ref{fig_ext:atlas_example_2}. For each source of the `GLMP
sample' a chart was put together as a mosaic of 4 images containing
the optical image taken from the DSS2 in the upper left panel, and the
J, H and K images from the 2MASS in the upper right, lower left and
lower right panels, respectively. The size of the field shown in each
band is 4.6\arcmin\,$\times$\,4.6\arcmin. For each source we marked
the position of the proposed optical/near infrared counterpart with a
circle in each of the available frames. In those cases where an
optical/near-infrared counterpart was not found the circle was drawn
at the position where the source should be located according to the
best astrometric coordinates available (IRAS or MSX). The complete
atlas can be accessed electronically at: http://www.edpsciences.org

   \section{\label{prop}Near-infrared properties}

\subsection{Near-infrared brightness distribution}
The typical near-infrared counterpart of objects from the `GLMP sample'
is very red (H--K$>$2).
75\%
of the GLMP sources were not detected in the J-band, 66\%
were not detected in the H band, and 37\%
were not detected even in the K band.

\begin{figure*}
\begin{center}
        \includegraphics[width=13cm]{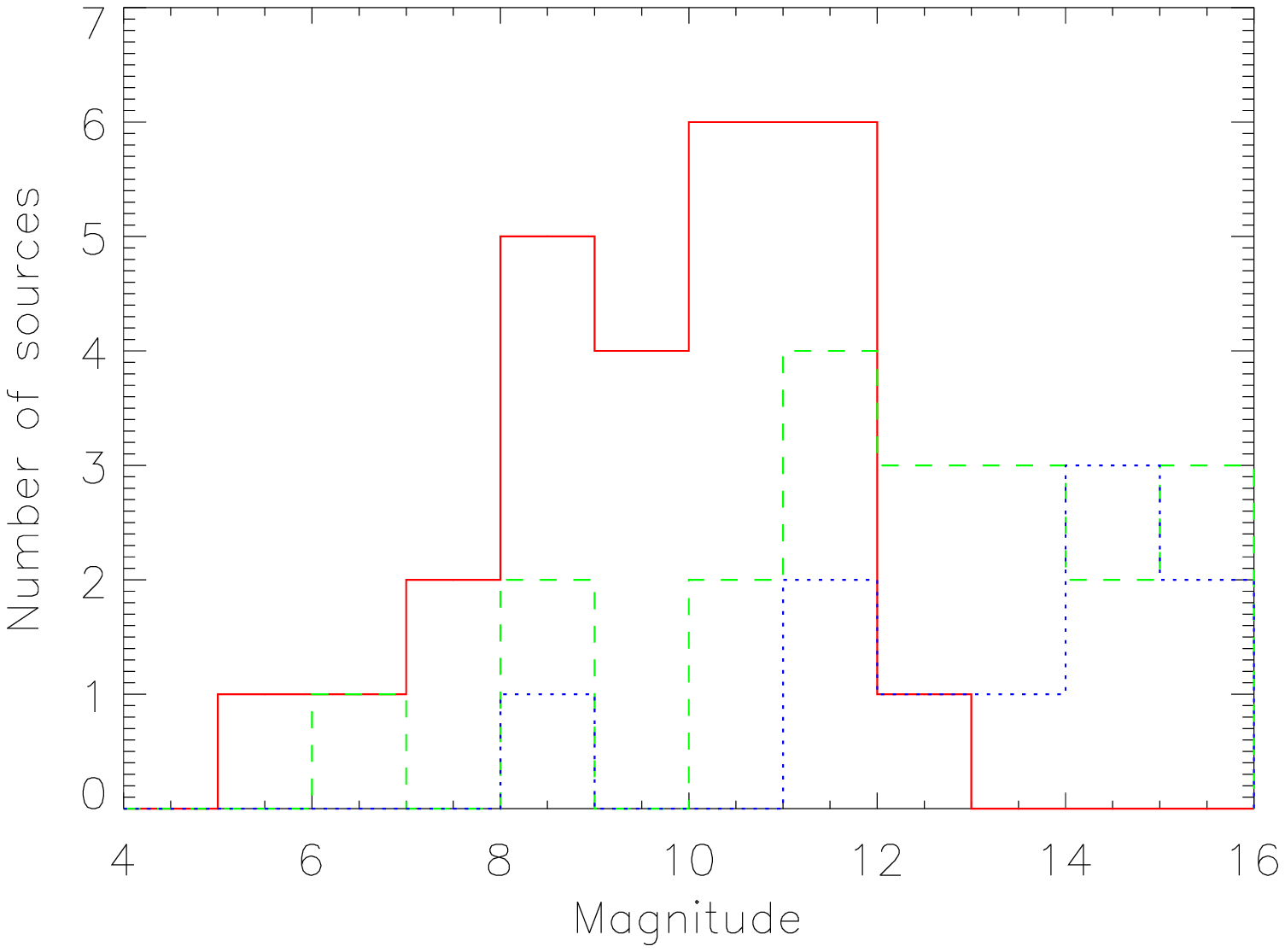} 
        \caption[Near-IR magnitude distribution of the sources
        included in the `GLMP sample'.]
        {\label{fig_ext:NIR_histogram_2}\,\ J-band ({\it dotted
        line}), H-band ({\it dashed line}), and K band ({\it solid
        line}) magnitude distribution of the sources included in the
        `GLMP sample'.}
\end{center}
\end{figure*}

Figure \ref{fig_ext:NIR_histogram_2} shows the J, H and K magnitude
distribution of all the GLMP sources with near-infrared counterparts
and photometric measurements in at least one band. Most of the sources
observed have faint near-infrared counterparts with the magnitude
distribution peaking at $\approx$ 10\Mag\, in K, $\approx$ 12\Mag\, in
H and $\geq$ 14\Mag\, in J. The distributions are not strongly peaked
and are clearly limited by the 2MASS sensitivity limit, especially in
the J band.

As it was discussed in Paper\,I for the sources in the `Arecibo
sample', the observed magnitude distribution can in principle be
attributed to three main reasons:

i) the different intrinsic luminosity of the sources in the sample,

ii) the different apparent brightness expected from sources located at
a wide range of distances, and

iii) the differences in optical thickness of the CSEs.


Compared to the magnitude distribution of the `Arecibo sample'
(Figure\,5 of Paper\,I) peaking at $\approx$ 6\Mag\, in K, $\approx$
6.5\Mag\, in H and $\geq$ 7.5\Mag\, in J, the mean near-infrared
brightness of the `GLMP sample' is several magnitudes fainter. The
sources in the `GLMP sample' are not expected to be intrinsically less
luminous than the ones in the `Arecibo sample' (actually we would
expect the contrary, since they are suspected to represent a higher
mass population). Thus, unless the sources in the `GLMP sample' are
preferentially found at larger distances than those in the `Arecibo
sample', the most probable explanation for the systematic fainter
brightnesses are the greater optical thickness of their CSEs, in
comparison with the optically thinner shells of the bluer Arecibo
sources.

\subsection{\label{ext_secJ-HvsH-K}Near infrared colours} 


\begin{figure*}
\begin{center}
   \resizebox{\hsize}{!}{.\includegraphics{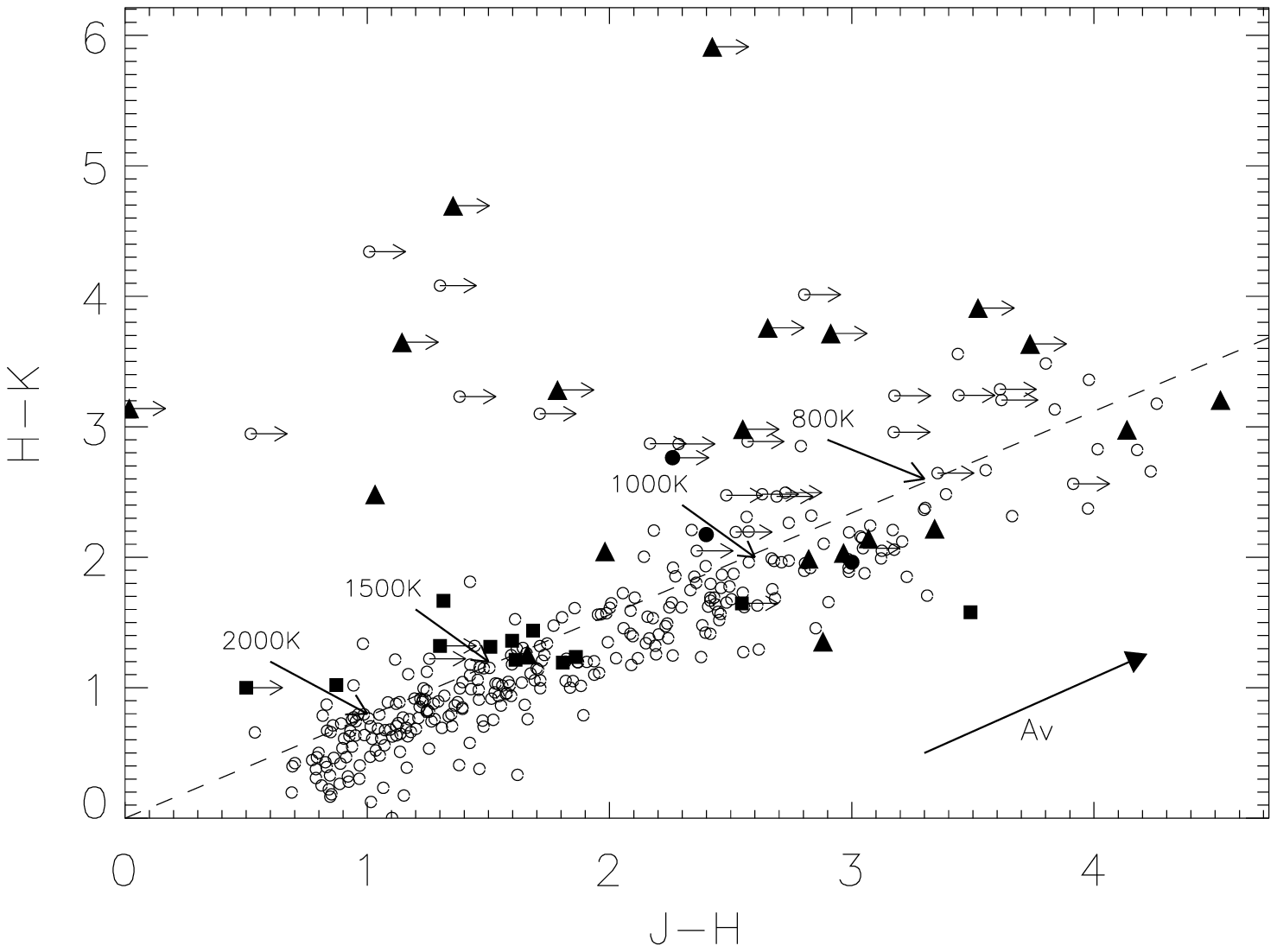}}
    \caption[J--H\,vs.\,H--K colour-colour diagram for the `GLMP
    sample']{\label{fig_ext:NIR_colour_2}\,\, Near-infrared
    J--H\,vs.\,H--K colour-colour diagram of the sources belonging to
    both the `red subsample' (filled triangles) and the `blue
    subsample' (filled squares) with H- and K-band detections
    (N\,=\,32). For comparison, the near-infrared colours of the
    sources in the `Arecibo sample' are also shown (open circles;
    filled circles for objects in common). J--H lower limits are
    indicated by horizontal arrows. The location of black-bodies of
    different temperatures and the reddening corresponding to
    A$_V$\,=\,10\Mag\, are also indicated.}
\end{center}
\end{figure*}


The near-infrared colours of the `GLMP sample' are shown as a
J$-$H\,vs. H$-$K colour-colour diagram in Figure
\ref{fig_ext:NIR_colour_2}. Included are all sources belonging to both
the `red subsample' (filled triangles) and the `blue subsample'
(filled squares) with at least data available in the H and K bands.
For objects not detected in J, lower limits for the J$-$H colours were
calculated and these are indicated by horizontal arrows.  The objects
with deblending problems have not been included. For comparison, we
have also plotted the colours of the sources in the `Arecibo sample'.

In general, the `GLMP sample' follows the correlation shown by the
more standard OH/IR stars of the `Arecibo sample', although the two
subsamples are also clearly separated in this diagram. The
distribution of the `red subsample' is restricted at the red end by
the limited sensitivity of 2MASS, especially in the J
band. Extrapolating the correlation between the colours, the reddest
object (IRAS\,11438--6330) with H--K\,=\,5.91 should have an expected
colour J--H\,$\approx$\,8 well beyond the limit of
J$-$H\,$\approx$\,4.5 following from Figure
\ref{fig_ext:NIR_colour_2}.

In Paper\,I we have shown that on average the redder an object is in
the far-infrared, the redder is its near-infrared counterpart (see
Figure\,8 of Paper\,I). The Arecibo sources with IRAS colours
satisfying the selection criteria imposed for the `GLMP sample'
(filled circles in Fig. \ref{fig_ext:NIR_colour_2}) are all located in
the red part of the near-infrared colour-colour
distribution. Actually, the near-infrared colours of the GLMP sources
overlap with the red end of the Arecibo J$-$H\,vs. H$-$K colour-colour
distribution. There are two outliers: IRAS\,$18299-1705$, which is
discussed below, and IRAS\,17171--2955, with H--K\,$\approx$\,2.5 and
J--H\,$\approx$\,1.0.  The 2MASS photometry of this source in the J-
and H-band is however biased by a nearby star that probably has
contaminated the background estimation. This makes its near-infrared
colours unreliable.  Except for IRAS\,$18299-1705$, the near-infrared
colours of the `red subsample' therefore corroborate the results
obtained from the `Arecibo sample' that the objects with optically
thicker CSEs reveal themselves in the far-infrared as well as in the
near-infrared by their redder colours.

\subsection{IRAS\,$18299-1705$: a post-AGB candidate?}

IRAS\,$18299-1705$ (J--H\,$\approx$\,1.7; H--K\,$\approx$\,1.3;
K=5.31) is located to the left of the dividing line in Figure
\ref{fig_ext:KvsH-K} and is therefore part of the `red subsample'.
However, in the J$-$H\,vs. H$-$K diagram it has very blue colours,
different from the rest of the `red subsample', and it has the
brightest near-infrared counterpart of the `GLMP sample'.  Instead of
the typical double-peaked maser profile of regular OH/IR stars, this
source shows a single peak 1612\,MHz OH maser (te Lintel Hekkert
\cite{teLintelHekkert91}). Its LRS spectrum does not show the very
strong 9.7\mic\ absorption feature characteristic of optically thick
CSEs, but a red continuum with either a weak 9.7\mic\ absorption
feature or the 11.3\mic\ PAH feature in emission (Kwok et
al. \cite{Kwok97}). This source is therefore peculiar.

Lewis et al. (\cite{Lewis04}) analysed the 2MASS counterparts of a
third of the `Arecibo sample' of OH/IR stars. They claimed that most
of the Arecibo sources with red far-infrared ([12]--[25]\,$\ga$\,0.0)
colour presented blue near-infrared colours in the J--K vs. H--K
diagram, similar to those of our `blue subsample'. They interpreted
these colours as the result of a drastic decrease of the mass-loss
rates from the central star leading to a decrease of the dust opacity
in the near environment of the central star. This allows the central
star to reappear in the near-infrared and to dominate the colours.
Since these objects have strongly decreased their mass-loss rate, the
authors inferred that these objects have detached shells.

IRAS\,$18299-1705$ is a good case for an AGB star with a detached
shell, and may have developed into the post-AGB phase. It is however a
single case in our sample, and we are therefore not able to confirm
the high incidence of detached shells among the AGB stars with very
red IRAS colours, as suggested by Lewis et al.
(\cite{Lewis04}). Photometric monitoring is required to confirm the
variability of IRAS\,$18299-1705$ inferred by the IRAS variability
index (VAR=74), which may contradict its post-AGB classifcation.

\subsection{The nature of the blue counterpart candidates}
The `blue subsample' is outstanding in the H--K\,vs.\,K (Fig. 
\ref{fig_ext:KvsH-K}) as well as in the J$-$H\,vs. H$-$K diagram 
 (Fig. \ref{fig_ext:NIR_colour_2}). Their blue near-infrared colours
are incompatible with the high optical thickness of their CSEs as
inferred from their IRAS colours. The only sources showing red
far-infrared colours and blue near-infrared colours are post-AGB
stars, e.g. stars with a detached shell. As post-AGB stars are no
longer variable, or at most with small amplitudes (Kwok
\cite{Kwok93}), we did not expect them to be part of the `GLMP
sample'.

The case of IRAS\,$18299-1705$ shows however that the IRAS variability
index might not be a reliable selection criterion to reject post-AGB
stars. There might be more of them among our `blue subsample'.  The
counterparts from the `blue subsample' differ however from
IRAS\,$18299-1705$ by their faintness (K$>$9 mag), and because of the
higher surface density of faint infrared sources they might be simply
misidentified field stars. Counterpart candidates with H--K$<$1.0 were
already rejected for this reason (see Section \ref{constraints}).

The case is less clear for the redder sources. Among the three Arecibo
sources with near-infrared colours of the `blue subsample' (see
Fig. \ref{fig_ext:KvsH-K}), two are known to have peculiarities. One
is IRAS\,18455+0448, for which Lewis (\cite{Lewis02}) found that the
OH maser decreased continuously and finally disappeared over
$<$\,12\,years. He argues that the mass-loss rate has dropped and that
the star is now leaving the AGB. The other is IRAS\,19319+2214, which
shows a peculiar OH maser profile with several peaks (Lewis et
al. \cite{Lewis04}). In this case the star may have several OH masing
shells expanding with different velocities. Among the `blue subsample'
is also the well known variable OH/IR star OH 19.2--1.0
(IRAS\,18266--1239; J--H\,$\approx$\,1.3; H--K\,$\approx$\,1.7),
showing a multiple-peak OH spectrum, which has been explained by a
bipolar structure of the CSE (Chapman \cite{Chapman88}). The
peculiarities of the OH masers in these three stars may indicate that
their mass-loss process (e.g. mass-loss rate or geometry of the
outflow) is currently experiencing fast changes, which might be
responsible for the relatively blue colours of their near-infrared
counterparts. The three cases show that the faint counterparts with
H--K\,$>$\,1 cannot be all misidentifications.  Instead, the deviating
near-infrared colours may point to AGB stars in which the mass loss
rates have recently varied drastically over time intervals comparable
to the dynamical time scale of their shells.

We conclude therefore that the nine remaining objects from the `blue
subsample' are a mixture of misidentifications and peculiar AGB stars,
requiring confirmation of their near-infrared counterparts by further
observations.

\subsection{Near- and far-infrared properties}


\begin{figure*}
\begin{center}  
   \resizebox{\hsize}{!}{.\includegraphics{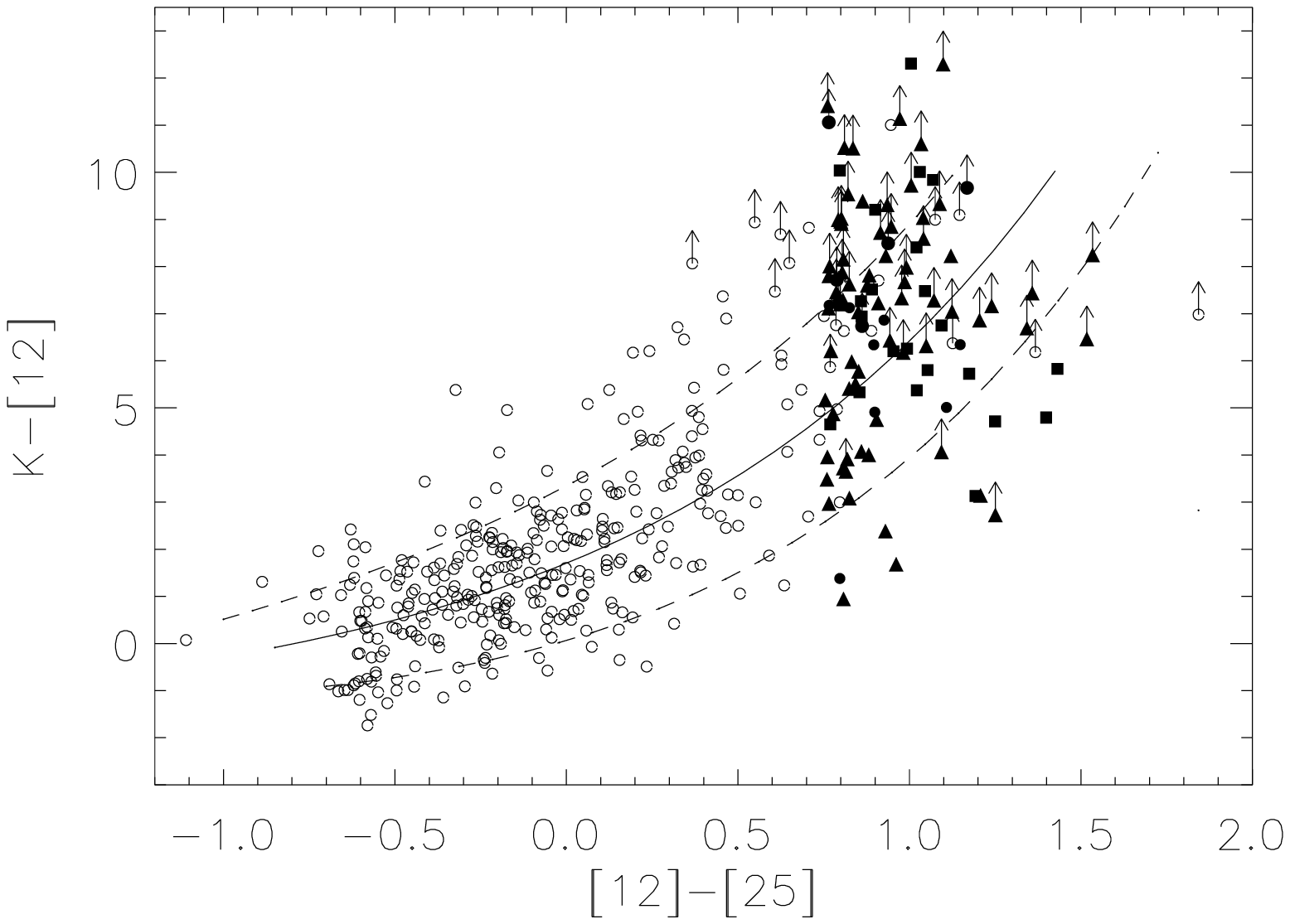}}
    \caption[Near/far-IR colour-colour diagram for the `GLMP
    sample']{\label{fig_ext:N-MIR_colour_2}\,\,
    K$-$[12]\,vs.\,[12]$-$[25] colour-colour diagram of all GLMP 
    stars with K band photometry (filled triangles; filled squares for
    those belonging the `blue subsample') together with those of the
    `Arecibo sample' (open circles; filled circles for objects in
    common). Upper limits in K were converted to lower limits in
    K$-$[12] colour (arrows). The continuous line corresponds to the
    best fit to the observed data (see text). The dashed lines
    correspond to the dispersion expected from the variability of the
    sources in the K and at 12\,$\mu$m band.}
\end{center}
\end{figure*}


A way to visualize the near- and far-infrared properties of the
sources in the GLMP sample is to study their distribution in the
K$-$[12]\,vs.\,[12]$-$[25] colour-colour diagram (Figure
\ref{fig_ext:N-MIR_colour_2}). In the blue part of the distribution 
the K$-$[12] colour provides information on the relative contribution
of the near-infrared emission, dominated by the central star and by
the hot dust surrounding it, and of the far-infrared emission, mainly
coming from the cool dust in the circumstellar shell, to the overall
spectral energy distribution. In the red part the K$-$[12] colour is
fully dominated by the outer cool CSE. As we have shown in Paper\,I, the
correlation of the K$-$[12] with the IRAS [12]$-$[25] colour shown by
the Arecibo OH/IR stars could be interpreted as an additional
indicator of the optical thickness of the CSE for a given source. In
Figure \ref{fig_ext:N-MIR_colour_2} we have plotted the GLMP OH/IR
stars together with those of the `Arecibo sample'.  For those sources
with no photometric measurement in K, upper limits were converted to
lower limits in the K$-$[12] colour. K$-$[12] colour was taken as
$-$2.5\,log$\frac{F_{\nu}(K)}{F_{\nu}(12)}$, adopting a zero-magnitude
flux in the K-band of 667\,Jy (Cohen et al. \cite{Cohen03}).

The scatter in Fig. \ref{fig_ext:N-MIR_colour_2} is mostly attributed to
the expected strong variability of the sources and the
non-contemporaneous near- and far-infrared observations. The expected
dispersion due to variability is indicated by the dashed lines (see
Paper I). Variability is however not able to explain fully the scatter
seen in Fig. \ref{fig_ext:N-MIR_colour_2}. For example, at [12]$-$[25]
$\approx$ 0.8 the range in K$-$[12] colour is $\approx$ 10 mag, which
is approximately twice the range expected due to variability.

In general, the `GLMP sample' of OH/IR stars confirms the colour trend
shown by the `Arecibo sample' of OH/IR stars. Using both samples we
have fitted an exponential function to the observational data,
weighting each data point according to the associated errors in each
colour and omitting the lower limits. We obtained the following
equation:
\begin{center}
K$-$[12]\,=\,3.44\,e$^{0.86([12]-[25])}$\,$-$\,1.74\\
\end{center}

This equation represents a parametrization of the colours of the
`O-rich AGB sequence', and has been plotted in Figure
\ref{fig_ext:N-MIR_colour_2} with a solid line. 

In general, the position of the GLMP OH/IR stars in the
K$-$[12]\,vs.\,[12]$-$[25] colour-colour diagram is an extension
toward redder colours of the sequence of colours associated with the
Arecibo sources. The `blue' and the `red subsample' defined in the
near-infrared cannot be distinguished in this diagram. The large
scatter of the K$-$[12] colours inhibits its use as an identification
criterion of near-infrared counterparts.

   \section{\label{conc}Conclusions}

We have presented an atlas of optical/near-infrared finding charts and
near-infrared photometric data for 94 OH/IR stars taken from the GLMP
catalogue, considered to be an extension toward redder colours and
thicker shells of the sources in the `Arecibo sample'. For 59 sources
we successfully identified their near-infrared counterparts in the
2MASS-PSC and determined new positions with an accuracy of
$\approx$\,0.2\arcsec. The identifications were possible in many cases
with the help of improved positional information taken from the MSX6C
(accuracy $\approx$\,1.8\arcsec) as an intermediate step, and
confirmed through the analysis of their near- and far-infrared
colours. For about a third of the sample (N=34 sources) no
near-infrared counterpart could be found even at 2.2\mic. These stars
most likely have the CSEs with the highest obscuration achieved by AGB
stars. 

As expected by the selection criteria, the near- and far-infrared
properties of most of the identified GLMP sources are similar to those
of the redder objects in the `Arecibo sample' analyzed in Paper\,I.
Some near-infrared counterparts showed surprisingly blue colours.
Among them is IRAS\,$18299-1705$, which has also the brightest 2MASS
counterpart. This source likely has a detached shell and might be a
post-AGB star. The other counterparts with blue colours are a mixture
of misidentifications with field stars and peculiar case, in which the
blue colours might be due to a recent strong decrease of the mass loss
rate leading to a corresponding decrease of the optical thickness of
their circumstellar shells.

The high fraction of GLMP sources still unidentified in the
near-infrared and the near- and far-infrared colours of the majority
of the sources identified in the 2MASS confirm their classification as
oxygen-rich AGB stars highly obscured by optically thick CSEs. They
belong to the same population of heavily reddened AGB stars, as the
reddest OH/IR stars in the `Arecibo sample', although not in all of
them has OH maser emission been detected yet. 

The observational results presented here, together with those already
included in Paper\,I, will be discussed in detail in a broader
astrophysical context in a future paper of this series.

   \begin{acknowledgements}

This work has been supported by the Spanish Ministerio de Ciencia y
Tecnolog\'\i a through travel grants (AYA2003$-$09499). This research
has made use of the SIMBAD database, operated at CDS, Strasbourg,
France. We acknowledge also the use of the Digitized Sky Survey, based
on photographic data obtained using the UK Schmidt Telescope. The UK
Schmidt Telescope was operated by the Royal Observatory Edinburgh,
with funding from the UK Science and Engineering Research Council,
until 1988 June, and thereafter by the Anglo-Australian
Observatory. Original plate material is copyright (c) of the Royal
Observatory Edinburgh and the Anglo-Australian Observatory. The plates
were processed into the present compressed digital form with their
permission. The Digitized Sky Survey was produced at the Space
Telescope Science Institute under US Government grant NAG W-2166. This
publication makes also use of data products from the Two Micron All
Sky Survey, which is a joint project of the University of
Massachusetts and the Infrared Processing and Analysis
Center/California Institute of Technology, funded by the National
Aeronautics and Space Administration and the National Science
Foundation.

   \end{acknowledgements}





\begin{thebibliography}{}

\bibitem[1987]{Bedijn87} 
Bedijn, P.~J.\ 1987, A\&A, 186, 136

\bibitem[1988]{Beichman88}
Beichman, C.~A., Neugebauer, G., Habing, J.~H., Clegg, P., \& Chester, T.~J.\ 1988, {\em IRAS Catalogs and Atlases, Volumen 1: Explanatory Supplement}, US Government Printing Office, Washington, DC

\bibitem[1989]{Bowers89}
Bowers, P.~F., \& Knapp, G.~R.\ 1989, ApJ, 347, 325

\bibitem[1988]{Chapman88}
Chapman, J.~M.\ 1988, \mnras, 230, 415 

\bibitem[1993]{Chengalur93}
Chengalur J.N., Lewis, B.M., Eder, J., Terzian, Y.\ 1993, ApJS, 89, 189

\bibitem[2003]{Cohen03}
Cohen, M., Wheaton, W.~A., \& Megeath, S.~T.\ 2003, AJ, 126, 1090

\bibitem[2005]{Cohen05}
Cohen, M., Parker, Q.~A., Chapman, J.\ 2005, \mnras, 357, 1189 

\bibitem[2003]{Cutri03}
Cutri, R.~M., Skrutskie, M.~F., van Dyk, S. et al.\ 2003, {\it 2MASS All-Sky Catalog of Point Sources}

\bibitem[2001]{Djorgovski01}
Djorgovski, S.~G., Gal, R.~R., de Carvalho, R.~R., et al.\ 2001, 199th American Astronomical Society Meeting, BAAS, 33, 1461 

\bibitem[1988]{Eder88}
Eder, J., Lewis, B.M., Terzian, Y.\ 1988, ApJSS, 66, 183

\bibitem[2003]{Egan03}
Egan, M.~P., Price, S.~D., \& Kraemer, K.~E.\ 2003, {The Midcourse Space Experiment (MSX) Point Source Catalog Version 2.3}

\bibitem[1983]{Engels83}
Engels, D., Kreysa, E., Schultz, G.~V., \& Sherwood, W.~A.\ 1983, A\&A, 124, 123

\bibitem[1992]{Garcia-Lario92}
Garc\'{\i}a-Lario, P.\ 1992, Ph.D. Thesis, Universidad de La Laguna, Tenerife (Spain)

\bibitem[1997]{Garcia-Lario97}
Garc\'{\i}a-Lario, P., Manchado, A., Pych, W., \& Pottasch, S.~R.\ 1997, A\&AS, 126, 479 

\bibitem[1989]{Gaylard89}
Gaylard, M.~J., West, M.~E., Whitelock, P.~A., \& Cohen, R.~J.\ 1989, MNRAS, 236, 247

\bibitem[1974]{Harvey74} 
Harvey, P.~M., Bechis, K.~P., Wilson, W.~J., \& Ball, J.~A.\ 1974, ApJS, 27, 331

\bibitem[1986]{Herman86}
Herman, J., Burger, J.~H., \& Penninx, W.~H.\ 1986, A\&A 167, 247

\bibitem[1993]{Hu93}
Hu, J.~Y., Slijkhuis, S., de Jong, T., \& Jiang, B.~W.\ 1993, A\&ASS, 100, 413

\bibitem[2005a]{Jimenez-Esteban05a} 
Jim\'{e}nez-Esteban, F.~M., Agudo-M\'{e}rida, L., Engels, D., \& Garc\'{\i}a-Lario, P.\ 2005a, A\&A, 431, 779

\bibitem[2005b]{Jimenez-Esteban05b} 
Jim\'{e}nez-Esteban, F.~M., Engels, D., \& Garc\'{\i}a-Lario, P.\ 2005b, in preparation

\bibitem[2005c]{Jimenez-Esteban05c} 
Jim\'{e}nez-Esteban, F.~M., Garc\'{\i}a-Lario, P. L., Manchado, A., \& Engels, D.\ 2005c, in preparation

\bibitem[1997]{Kwok97} 
Kwok, S., Volk, K., \& Bidelman, W.~P.\ 1997, \apjs, 112, 557

\bibitem[1993]{Kwok93} 
Kwok, S.\ 1993, \araa, 31, 63 

\bibitem[1995]{Lepine95}
Lepine, J.~R.~D., Ortiz, R., \& Epchtein, N.\ 1995, A\&A, 299, 453

\bibitem[2004]{Lewis04} 
Lewis, B.~M., Kopon, D.~A., \& Terzian, Y.\ 2004, AJ, 127, 501 

\bibitem[2002]{Lewis02}
Lewis, B.~M.\ 2002, \apj, 576, 445 

\bibitem[1992]{Lewis92} 
Lewis, B.~M.\ 1992, \apj, 396, 251

\bibitem[1990]{Lewis90}
Lewis, B.~M., Eder, J., Terzian, Y.\ 1990, ApJ, 362, 634

\bibitem[1984]{Olnon84}
Olnon, F.~M., Baud, B., Habing, H.~J., de Jong, T., Harris, S., \& Pottasch, S.~R.\ 1984, ApJ (Letters) 278, 41

\bibitem[2005]{PereaCalderon05}
Perea Calderon, J.V., Garc\'{\i}a-Lario, P., Jim\'{e}nez-Esteban, F.~M., \& Su\'{a}rez, O.\ 2005, in preparation

\bibitem[1991]{teLintelHekkert91}
te Lintel Hekkert, P., Caswell, J.~L., Habing, H.~J., Haynes, R.~F., Haynes, R.~F., \& Norris, R.~P.\ 1991, \aaps, 90, 327 

\bibitem[1995]{Wendker95}
Wendker, H.~J.\ 1995, \aaps, 109, 177 

\bibitem[1994]{Whitelock94}
Whitelock, P., Menzies, J., Feast, M., et al.\ 1994, MNRAS, 267, 711

\bibitem[1989]{Zijlstra89}
Zijlstra, A.~A., Te Lintel Hekkert, P., Pottasch, S.~R., Caswell, J.~L., Ratag, M., \& Habing, H.~J.\ 1989, \aap, 217, 157

   \end{thebibliography}
\end{document}